\documentclass[twocolumn,showpacs,aps,prd]{revtex4b5}

\usepackage{graphicx}
\usepackage{dcolumn}
\usepackage{amsmath}
\usepackage{epsfig}
\usepackage{hhline}

\input pubboard/babarsym
\def\Dp      {\ensuremath{D^+}}
\def\Dm      {\ensuremath{D^-}}
\def\Dstar   {\ensuremath{D^{*}}}
\def\Dstarp  {\ensuremath{D^{*+}}}
\def\Dstarm  {\ensuremath{D^{*-}}}

\newcommand{\etal}{{\it et al.}}

\newcommand{\BABARPubYear}    {01}
\newcommand{\BABARPubNumber}  {07}

\newcommand{\SLACPubNumber} {8909}

\newcommand{\bpjpsikp}{\ensuremath{B^+ \to \jpsi K^+}}
\newcommand{\bppsitwoskp}{\ensuremath{B^+ \to \psitwos K^+}}
\newcommand{\bpchiconekp}{\ensuremath{B^+ \to \chic{1} K^+}}
\newcommand{\bzjpsikstarz}{\ensuremath{B^0\to \jpsi K^{*0}}}
\newcommand{\bpjpsikstarp}{\ensuremath{B^+ \to \jpsi K^{*+}}}
\newcommand{\bzjpsiks}{\ensuremath{B^0 \to \jpsi\KS}}
\newcommand{\bzjpsikl}{\ensuremath{B^0 \to \jpsi\KL}}
\newcommand{\bzjpsikz}{\ensuremath{B^0 \to \jpsi K^{0}}}
\newcommand{\bzjpsipiz}{\ensuremath{B^0 \to \jpsi\piz}}
\newcommand{\bzpsitwoskz}{\ensuremath{B^0 \to \psitwos K^{0}}}
\newcommand{\bzpsitwosks}{\ensuremath{B^0 \to \psitwos\KS}}
\newcommand{\bzchiconeks}{\ensuremath{B^0 \to \chic{1}\KS}}
\newcommand{\bzchiconekz}{\ensuremath{B^0 \to \chic{1} K^{0}}}
\newcommand{\bzchiconekstarz}{\ensuremath{B^0 \to \chic{1}K^{*0}}}

\newcommand{\De} {\ensuremath{\Delta E}}
\newcommand{\DeKL} {\ensuremath{\Delta E_{\KL}}}

\newcommand{\absDe} {\ensuremath{|\Delta E|}}

\def\figurebox#1#2#3{%
    \def\arg{#3}%
    \ifx\arg\empty
    {\hfill\vbox{\hsize#2\hrule\hbox to #2{\vrule\hfill\vbox to #1{\hsize#2\vfill}\vrule}\hrule}\hfill}%
    \else
    {\hfill\epsfbox{#3}\hfill}%
    \fi}

\long\def\inst#1{\par\nobreak\kern 4pt\nobreak
    {\it #1}\par\vskip 10pt plus 3pt minus 3pt}

\begin{document}


\begin{flushleft}
\babar-PUB-\BABARPubYear/\BABARPubNumber\\
SLAC-PUB-\SLACPubNumber\\
\end{flushleft}

\title{\large \bf
Measurement of branching fractions for exclusive $B$ decays to charmonium
final states
\begin{center} 
\vskip 10mm
The \babar\ Collaboration
\end{center}
}

%
\author{B.~Aubert}
\author{D.~Boutigny}
\author{J.-M.~Gaillard}
\author{A.~Hicheur}
\author{Y.~Karyotakis}
\author{J.~P.~Lees}
\author{P.~Robbe}
\author{V.~Tisserand}
\affiliation{Laboratoire de Physique des Particules, F-74941 Annecy-le-Vieux, France }
\author{A.~Palano}
\affiliation{Universit\`a di Bari, Dipartimento di Fisica and INFN, I-70126 Bari, Italy }
\author{G.~P.~Chen}
\author{J.~C.~Chen}
\author{N.~D.~Qi}
\author{G.~Rong}
\author{P.~Wang}
\author{Y.~S.~Zhu}
\affiliation{Institute of High Energy Physics, Beijing 100039, China }
% ...
\collaboration{The \babar\ Collaboration}
\noaffiliation

\date{\today}

\begin{abstract}
We report branching fraction measurements for
 exclusive decays of charged and neutral $B$ mesons into 
two-body final states containing
a charmonium meson.  We use a sample of $22.72 \pm 0.36$ million \BB\ events 
collected  between October 1999 and October 2000 with the 
\babar\ detector at the PEP-II storage rings at the Stanford Linear 
Accelerator Center.
The charmonium mesons considered here are \jpsi, \psitwos, and 
\chic{1}, and the light meson in the decay is either a \kaon, \Kstar, or
\piz.  

\end{abstract}

\pacs{11.30.Er, 13.25.Hw}


\maketitle

\section{Introduction}
\label{sec:Introduction}

Decays of $B$ mesons to two-body final states containing a charmonium 
resonance 
(\jpsi, \psitwos, $\chicone$)
constitute a very sensitive laboratory for the study of electroweak 
transitions, 
as well as the dynamics of strong interactions in heavy meson systems. 
In particular, neutral $B$ decays to these final states are expected to exhibit
a significant \CP\ asymmetry, the magnitude of which is cleanly related to
standard model parameters~\cite{ref:sin2b}.

\begin{figure}
\begin{center}   
\epsfig{file=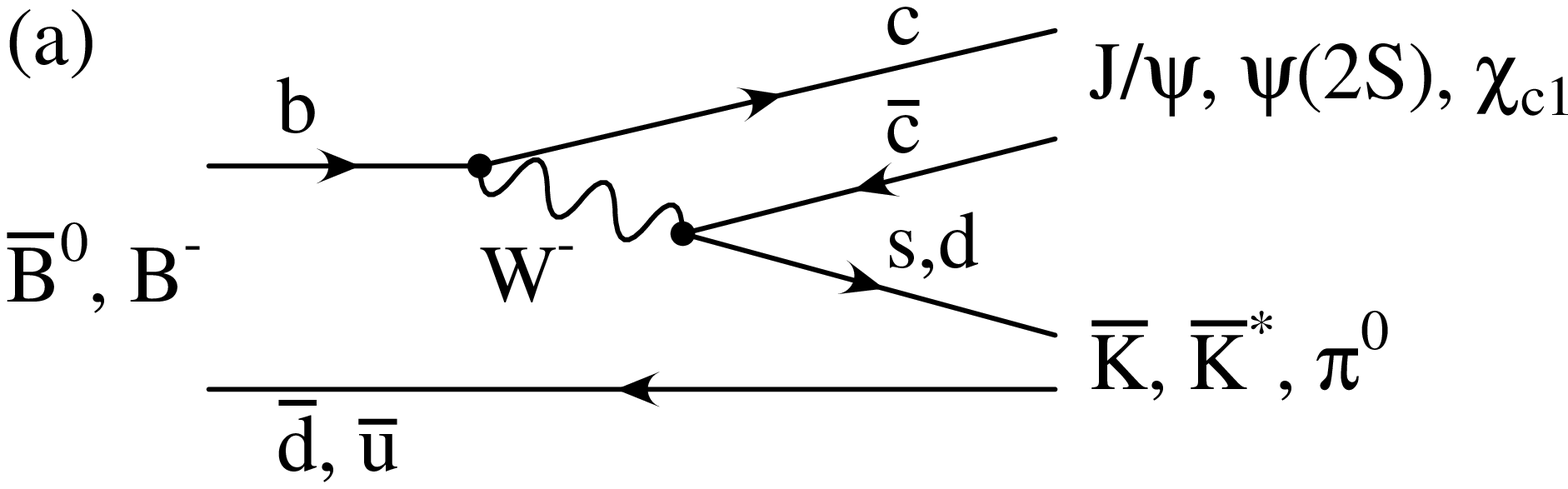,width=0.45\textwidth} 
\epsfig{file=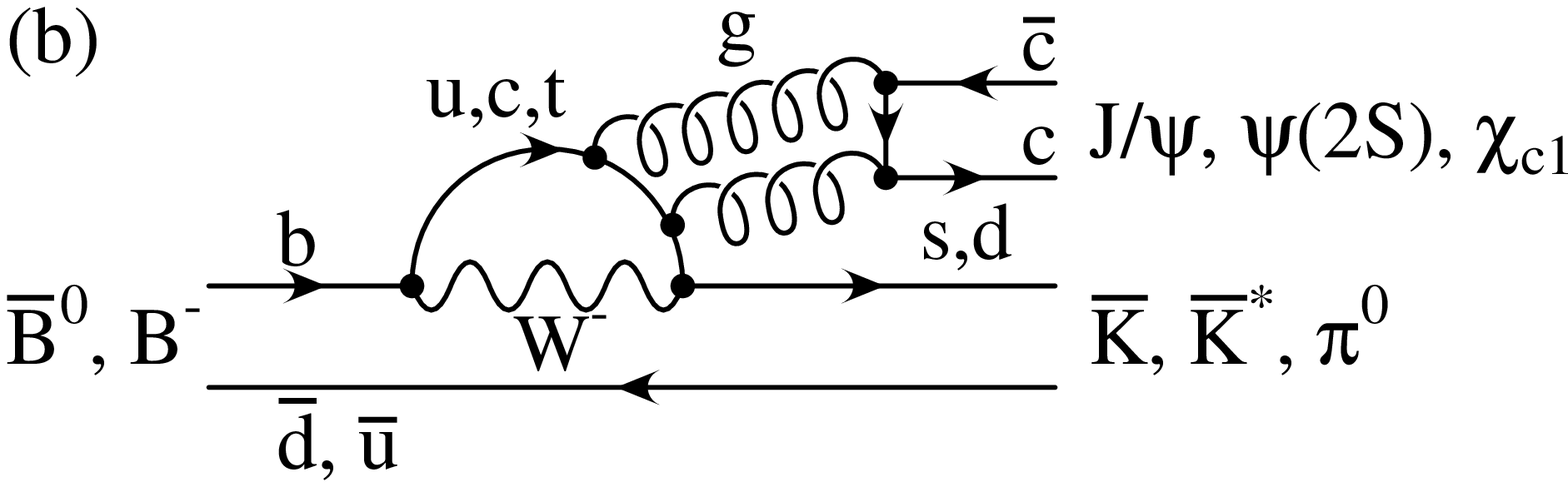,width=0.45\textwidth} 
\end{center}
\caption[decaydiagram]{Leading Feynman diagrams for the decays we consider. 
\label{fig:decaydiagram}}
\end{figure}

The tree level and leading penguin diagrams for the decay modes we consider
are shown in Fig.~\ref{fig:decaydiagram}. 
Due to the contributions of non-perturbative QCD interactions in the final
state, assumptions must be made in estimating the expected branching
fractions of these modes, and therefore these estimates have some degree
of model dependence.  A number of such estimates have appeared in the 
literature~\cite{WBS,ISW,nsrx,deand,aleks,ns97,Ciuchini,Keum,Yeh,
Martinelli,Cheng}.  The one model-independent element common to all of these
predictions is the requirement from isospin symmetry that the ratio of the 
charged to neutral partial widths should
be unity, and that this should hold separately for each light meson 
accompanying the charmonium meson in the final state.  

Here we report the measurement of branching fractions of $B$ mesons to a 
charmonium resonance accompanied by a kaon or \piz\ meson. The channels 
measured are listed in Table~\ref{listofchannels}.
Here and throughout this paper for each final state mentioned its charged 
conjugate is also implied. We reconstruct \jpsi\ decays to 
lepton pairs \ellell, where $\ell$ is either an electron or muon.

Our large data sample permits a measurement of these branching 
fractions with a precision superior to previous experiments.
The simultaneous measurement of a number of final states allows us to
determine ratios such as vector to pseudoscalar kaon and 
heavy to light charmonium states production. 
Many systematic errors cancel when these ratios are extracted from a single 
data set using very similar event selection criteria, further increasing the 
usefulness of our results for the validation or development of 
phenomenological models.

\begin{table}
\caption[tab:modes]{Branching fractions and decay modes considered in this 
paper.  We always
reconstruct the \jpsi\ in the \ellell\ decay mode.
\label{listofchannels}}
\begin{center}
\begin{tabular}{ll} \hline \hline
\multicolumn{1}{c}{Branching fraction} & \multicolumn{1}{c}{Secondary decay}                                    \\ 
\multicolumn{1}{c}{measured}  & \multicolumn{1}{c}{modes used} \\ \hline
\bzjpsikz        & \Kz\to\KS; \KS \to \pipi or \piz\piz      \\ 
                 & \Kz\to\KL  \\
\bpjpsikp        &  --                                         \\
\bzjpsikstarz    &  \Kstarz \to \Kp\pim or \KS\piz; \KS\to\pipi        \\
\bpjpsikstarp    &  \Kstarp \to \Kp\piz or \KS\pip; \KS\to\pipi         \\
\bzjpsipiz       &  --                                          \\ \hline
\bzpsitwosks     & \psitwos\to\ellell or \jpsi\pipi; \\
                 &       \multicolumn{1}{r}{\KS\to\pipi}  \\
\bppsitwoskp     & \psitwos\to\ellell or \psitwos\to\jpsi\pipi              \\ \hline
\bzchiconeks     & \chic{1}\to\jpsi\g ; \KS\to\pipi                        \\
\bpchiconekp     & \chic{1}\to\jpsi\g                                      \\ 
\bzchiconekstarz & \chic{1}\to\jpsi\g ; \Kstarz \to \Kp\pim                \\ \hline \hline
\end{tabular}
\end{center}
\end{table}

Another highly relevant input for the understanding of strong interactions 
in $B$ 
decays is the measurement of polarization in vector-vector final states, 
which is 
reported in another publication~\cite{AngAnalysis}.
Finally, the branching fraction of 
$B \rightarrow \jpsi \pi^{+}$ 
is measured using a specific analysis method, reported in~\cite{JpsiPitoJpsiK} 
.

\section{The \babar\ detector}
\label{sec:babar}

The \babar\ detector is located at the \pep2\ \epem\ storage rings
operating at the Stanford Linear Accelerator Center.
At PEP-II, 9.0\gev\ electrons collide with 3.1\gev\ positrons to produce
 a center-of-mass energy of 10.58\gev, the mass of the \FourS\ resonance.

The \babar\ detector is described elsewhere~\cite{ref:babar}; here we give 
only a brief overview.
Surrounding the interaction point is a 5-layer double-sided
silicon vertex tracker (SVT) 
which gives precision spatial information for
all charged particles, and also measures their energy loss (\dedx).
The SVT is the primary detection device for low momentum
charged particles. Outside the SVT, a 40-layer drift chamber (DCH) 
provides measurements of the transverse momenta \pt\ of charged particles
with respect to the beam direction. 
The resolution of the \pt\ measurement for tracks with momenta above 1 \gevc 
is parameterized as:
\begin{equation}
\frac{\sigma(\pt)}{ \pt} = 0.13\pt(\gevc)\% + 0.45\%.
\end{equation}
The drift chamber also measures
\dedx\ with a precision of 7.5\%.
Beyond the outer radius of the DCH is a detector of internally reflected 
Cherenkov radiation (DIRC) which is used primarily for charged hadron
identification. The  detector consists of quartz bars in
which Cherenkov light is produced as relativistic charged particles traverse
the material. The light is internally reflected along the length of the bar into a water-filled 
stand-off box mounted on the rear of the detector.  The Cherenkov rings expand in the stand-off box and 
are measured with an array of photomultiplier tubes mounted on its outer surface. A CsI(Tl) crystal 
electromagnetic calorimeter (EMC) is used to 
detect photons and neutral hadrons, 
as well as to identify electrons.  The resolution of the calorimeter is 
parameterized as:
\begin{equation}
\frac{\sigma(E)}{E} = \frac{2.3\%}{(E(\gev))^{\frac{1}{4}}} \oplus 1.9\%.
\end{equation}
The EMC is surrounded by a superconducting 
solenoid that produces a 1.5-T magnetic field. The instrumented flux return
(IFR) consists of multiple layers of resistive plate chambers (RPC) 
interleaved with the flux return iron. In addition to the planar RPC layers
in the flux return, there is an additional cylindrical layer just outside of the
EMC.  The IFR is
used in the identification of muons and neutral hadrons. 

Data acquisition is triggered with a two-level system.  The first level 
(Level 1) monitors trigger information from
the DCH and EMC, and generates a trigger upon detection of track or cluster 
candidates.  The second level (Level 3) retains events in which 
the track candidates point back 
to the beam interaction region (L3 DCH trigger),  or EMC clusters candidates
remain
 after the suppression of hits which have less energy
 than a minimum ionizing particle or are uncorrelated in time with the rest 
of the event (L3 EMC trigger).  Over 99.9\% of \BB\ events pass either the 
L3 DCH or L3 EMC trigger.
  A fraction of all events that pass
the Level 1 trigger are passed through Level 3 to allow monitoring of the Level
3 trigger performance.

\section {Data Sample}
     
The data used in these analyses 
were collected
between October 1999 and October 2000 and 
correspond to an integrated 
luminosity of 20.7\invfb\ taken 
on the $\FourS$ and 2.6\invfb\ taken off-resonance at an energy 0.04 GeV 
lower than the peak, which is below the threshold for \BB\ production. 
The data set contains  $22.72 \pm 0.36$ million \BB\ events.

\section {Coordinate System and Reference Frames}

We use a right-handed coordinate system with the $z$ axis along the electron
beam direction and $y$ axis upwards, with origin at the nominal beam
interaction point.  Unless otherwise stated, kinematic quantities are 
calculated in the rest frame of the detector.  The other reference frame we 
commonly use
is the center of mass of the colliding electrons and positrons, which 
we will call the center-of-mass frame.

\section{Particle reconstruction}

\begin{table*}
\caption{
\label{tab:eselec}
Summary of electron identification criteria.  Variables used are: \dedx, the
energy loss measured in the DCH; 
$E/p$, the ratio of the EMC cluster energy to the
momentum measured in the tracking spectrometer; 
$N_{\rm crys}$, the number of EMC crystals forming the cluster;
LAT, the lateral energy distribution \cite{ref:lat} of the EMC cluster; 
$A_{42}$, one of the  Zernike moments \cite{ref:zernike} of the EMC cluster;
 and $\theta_C$, the Cherenkov angle measured in the DIRC.  
In addition, the fraction of electrons in inclusive \jpsi\ 
events that pass each set of criteria is shown, along with the fraction
of pions with momentum above 1 \gevc\ that pass the selection requirements.
}
\begin{center}
\begin{tabular}{lcccc}
\hline \hline
         & DCH-only & Loose & Tight & Very tight \\ \hline
\dedx\ (measured-expected) &  -2 to +4 $\sigma_{\rm meas}$  & -3 to +7 $\sigma_{\rm meas}$  & -3 to +7 $\sigma_{\rm meas}$  & -2 to +4 $\sigma_{\rm meas}$  \\
$E/p$    &    --     &  0.65 - 5.0  &  0.75 - 1.3 & 0.89 - 1.2 \\
$N_{\rm crys}$ & --  & $>3$  & $>3$  & $>3$        \\
LAT      &  --      & --   & 0.0 - 0.6 & 0.1 - 0.6 \\
$A_{42}$ &  --      & --   & --  & $< 0.11$ \\
$\theta_C$ (measured-expected)  & --  & -- & -- & -3 to +3 $\sigma_{\rm meas}$ \\ \hline
Efficiency (\%)     & 94.9  &  97.2  & 95.4 & 88.2 \\
$\pi$ misID (\%)    & 21.6  &  4.8   &  1.2 & 0.1 \\ \hline \hline
 \end{tabular}
\end{center}
\end{table*}

\begin{table*}
\caption{
\label{tab:muselec}
Summary of muon identification criteria.  Variables used are: $E_{\rm EMC}$, 
the
energy deposited by the muon candidate in the EMC (this requirement is only 
applied for
tracks within the fiducial coverage of the EMC); $N_{\rm layers}$, the 
number of IFR layers with hits;
$N_{\lambda}$, the number of nuclear interaction lengths traversed;
$\vert N_{\lambda}-N_{\lambda}\rm{(exp)}\vert$, the difference between the
number of nuclear interaction lengths traversed and the expectation for 
a muon of the measured momentum;
$\langle N_{\rm hit}\rangle$, the average number of hits per IFR layer;
RMS$_{\rm hit}$, the RMS of the distribution of the number of hits on each 
layer;
$f_{\rm hit}$, the fraction of layers between the innermost and outermost
hit layers that also have hits (this requirement is only applied in the 
region covered partly or
entirely by the endcap IFR system, $0.3 < \theta < 1.0$);
$\chi^2_{\rm IFR}$, the $\chi^2$ of the track in the IFR;
and $\chi^2_{\rm match}$, the $\chi^2$ of the match between the IFR track and
the track from the central detector.  In addition, the fraction of muons in 
inclusive \jpsi\  events that pass each set of criteria is shown, along with 
the fraction of pions with momentum above 1 \gevc\ that pass the 
selection requirements.
}
\begin{center}
\begin{tabular}{lccccc}
\hline\hline
         & MIP & Very Loose & Loose  & Tight  & Very tight \\ \hline
$E_{\rm EMC}$  (\gev)  &   $< 0.5$    &   $< 0.5$  &  $< 0.5$ & $0.05 - 0.4$ & $0.05 - 0.4$ \\
$N_{\rm layers}$ & --  & $>1$  & $>1$  & $>1$ & $>1$        \\
$N_{\lambda}$    &  --       & $>2$   & $>2$ & $>2.2$ & $>2.2$ \\
$\vert N_{\lambda}-N_{\lambda}\rm{(exp)}\vert$ &  --       & $<2.5$  & 
$<2.0$     & $<1$ & $<0.8$ \\
$\langle N_{\rm hit}\rangle$ & --  & $<10$ & $<10$ & $<8$ & $<8$ \\
RMS$_{\rm hit}$ & -- & $< 6$ &  $< 6$ & $<4$ & $<4$ \\
$f_{\rm hit}$ & -- & $>0.1$ & $>0.2$ & $>0.3$ & $>0.34$ \\
$\chi^2_{\rm IFR}$ & -- & -- & $<4 \times N_{\rm layers}$ & $<3 \times N_{\rm layers}$ & $<3 \times N_{\rm layers}$ \\
$\chi^2_{\rm match}$ & -- & -- & $< 7\times N_{\rm layers}$ & $< 5\times N_{\rm layers}$ & $< 5\times N_{\rm layers}$ \\ \hline
Efficiency (\%)     & 99.6  &  92.2  & 86.2 & 70.3 & 67.0 \\
$\pi$ misID (\%)    & 57.9  &  14.5   &  7.0 & 2.4 & 2.1 \\ \hline\hline
 \end{tabular}
\end{center}
\end{table*}

The reconstruction of exclusive $B$ decays begins with identifying candidates for
the decay products.  Charged particles are reconstructed as
tracks in the SVT and/or DCH.  Leptons and kaons are identified with 
information
from the DCH, the EMC (for electrons) the IFR (for muons), and the 
DIRC (for kaons).
Photons are identified based on their energy 
deposition in the EMC, and \KL\ are identified from either energy 
deposition in the EMC or a shower in the IFR.

\subsection{Track Selection}
In general, tracks used in this analysis are required to include at least
12 DCH hits to ensure that their momenta and
\dedx\ are well measured.  In addition, tracks are required to have 
\pt $> 100$ \mevc, and to point back to the nominal 
interaction point within 1.5 cm in $xy$ and 3 cm in $z$.
Roughly 95\% of the solid angle about the interaction point in the
center-of-mass frame is covered by 12 or more DCH layers.

We make exceptions to this requirement for two types of particles:  
pions from \KS, which do not 
originate at the nominal 
interaction point, and pions from \mbox{\psitwos\to\jpsi\pipi}, 
which frequently do not have sufficient
transverse momenta to traverse 12 layers of the DCH.  Any track 
found in the DCH or SVT is used in reconstructing these particles.

\subsection{EMC cluster reconstruction}
The energy deposited in contiguous crystals of the EMC is summed into a
cluster.  The distribution of energy among the crystals is used to 
discriminate between clusters arising from electromagnetic and hadronic showers.
 The variables used to describe this distribution are the lateral energy
 (LAT)~\cite{ref:lat} 
and the Zernike moments $A_{mn}$~\cite{ref:zernike}. 
LAT is a measure of the radial energy profile of the cluster; the 
Zernike moment $A_{42}$ measures the asymmetry of the cluster about its
maximum. 
Electromagnetic
showers have LAT peaked at about 0.25 and $A_{42}$ close to zero, while
showers from hadrons have a broader distribution in LAT,
and extend to larger values of $A_{42}$.

\subsection{Photon Candidate Selection}
\label{sec:photon}
 Photons are identified as EMC clusters that do not have a spatial match
 with a
charged track, and that have a minimum energy of 30\mev.  To reject
clusters arising from noise hits, LAT is required to be less than 0.8.

\subsection{Electron and Muon Identification}

We derive substantial background rejection from the positive identification
of electrons and muons within the sample of charged tracks.
For electrons, the variables that distinguish signal from 
background include LAT and $A_{42}$, the ratio of energy measured in 
the EMC to momentum measured in the tracking spectrometer ($E/p$), 
\dedx\ measured in the DCH, and the 
Cherenkov angle $\theta_C$ measured in the DIRC.

 For identifying muons, the presence of an energy
deposition consistent with a minimum ionizing particle in the EMC, and 
the details of the
distribution of hits in the IFR are used.  In particular, the number
of interaction lengths traversed in the IFR $N_\lambda$ must be consistent with
expectations for a muon, both the average and variance of the number
of hits per layer must be small, and the fit of a track to the hits must
have low $\chi^2$, both within the IFR ($\chi^2_{\rm IFR}$) and in the 
match between the IFR and central detector track ($\chi^2_{\rm match}$).

Since the optimal tradeoff between efficient selection and suppression of 
backgrounds varies between decay modes, there are several sets of criteria used
to select leptons.  These are defined in
Table~\ref{tab:eselec} for electrons and Table~\ref{tab:muselec} for muons.
In addition to these criteria, we also restrict the lepton selection to a 
fiducial region within which the efficiency is well-known
from control samples, and the material in the detector is accurately modelled 
in the Monte Carlo.  The accepted range in polar angle $\theta$ is 
$0.410 < \theta < 2.409 \rad$ for 
electrons and $0.30 < \theta < 2.70 \rad$ for muons.  This corresponds to a 
coverage of 84\% of the solid angle in the center-of-mass 
frame for electrons, and
92\% for muons.

To increase the efficiency of the event selection, electron candidate tracks 
are combined with photon candidates to recover some of the energy lost
through bremsstrahlung.  In addition to the photon selection criteria 
listed above, photons
used in bremsstrahlung recovery are required to have $A_{42} < 0.25$.  They are
also required to be within 35 \mrad\ in $\theta$ from the track, and to have
azimuthal angle $\phi$ intermediate between the initial track direction and the centroid of 
the EMC cluster arising from the track. The initial track direction is 
estimated by subtracting 50 \mrad\ opposite to the bend direction from the
$\phi$ of the fitted track measured at the origin.  The procedure increases the 
efficiency for reconstructing charmonium decays to \epem\ by about 30\%.

\subsection{ \KL Candidate Selection}
We identify neutral hadrons through the presence of an energy 
deposition in the EMC or a cluster in the IFR.  Neutral hadrons must be 
spatially separated from all tracks in the event. In reconstructing the 
decay \bzjpsikl\ neutral hadrons are taken as \KL\ candidates, with
requirements specifically tailored for this mode.

Only the measured direction of the neutral hadron is 
used for \KL\ reconstruction, as its energy is poorly measured.
  The direction of the \KL\ candidate is 
defined by the line joining the vertex of the \jpsi candidate and the 
centroid of the 
EMC or IFR cluster.

For a \KL\ to reach the IFR it must traverse the EMC material, which 
amounts to 
approximately one nuclear interaction length. As a consequence, half 
of the \KL\ mesons undergo detectable interactions in the EMC.  
 We consider EMC clusters with energy in the 
0.2 - 2.0 GeV range.  Most clusters arising from \KL\ interactions
have energy below the
upper bound; below the lower bound the contamination from noise becomes
significant.  All such 
EMC clusters which are spatially separated from a track are considered 
as \KL\ candidates, except those that combined with another neutral 
cluster give an invariant mass compatible with a \piz.

About 60\% of \KL\ mesons from \bzjpsikl\ leave 
a detectable signal in the IFR.  
We select \KL\ candidates in the IFR starting with clusters of hits not
spatially matched to a track.
IFR clusters with hits only in the outer layers of the forward endcap are 
rejected to reduce the contribution from beam backgrounds.

\label{sec:PartReco}
\section{Event Selection and $B$ Meson Counting}
A determination of $B$ meson branching fractions depends upon an accurate
measurement
of the number of $B$ mesons in the data sample.  
We find the number of \BB\ pairs by comparing the rate of multihadron events
in data taken on the \FourS\ resonance to that in data taken off-resonance.
  The \BB\ purity of the sample is enhanced by requiring the 
events to pass the following selection criteria, in which all tracks 
(including those that do not satisfy our usual selection requirements) in the
fiducial region $0.410 < \theta < 2.54 \rad$ and all neutral clusters with 
energy
greater than 30 \mev\ in the region $0.410 < \theta < 2.409 \rad$ are 
considered:

\begin{itemize}
\item The event must satisfy either the L3 DCH or L3 EMC trigger.

\item There must be at least three tracks that satisfy the standard selection
requirements
in the fiducial region.
\item The ratio of the second to the zeroth Fox-Wolfram moment~\cite{FoxWolf}
must be less than 0.5.
\item The event vertex is calculated by an iterative procedure that begins
by considering every track in the event, and then discards those that
contribute a large $\chi^2$ to the fit (these are presumed to arise from the
decay of long-lived particles) until the vertex fit is stable.  This vertex 
 must be 
within 0.5 cm of the beam spot center in $xy$ and within 6 cm in $z$.  
The beam
spot has an RMS width of about 120 $\mu$m in $x$, 5.9 $\mu$m in $y$, and 
0.9 cm in $z$.  The point of closest approach of a high-momentum 
track to the beam spot
is measured with a resolution of 23 $\mu$m in $x$ and $y$, and  29 $\mu$m in
$z$, as determined with dimuon events.

\item The total energy of charged and neutral particles
 is required to be 
greater than 4.5\gev.
\end{itemize}

These requirements are $95.4 \pm 1.4$\% efficient for \BB\ events, as
 estimated from a Monte Carlo simulation.  All events used in the
branching fraction analyses are required to pass this selection.

\section{Meson Candidate Selection}
\begin{table}
\caption{Summary of observed invariant mass or mass difference $\Delta m$
widths for all intermediate mesons considered 
in this paper.  For most mesons the width is dominated by experimental 
resolution, and the value reported in the table is the width $\sigma$
from a 
Gaussian fit to the data.  For the \Kstar\ modes the natural width of the
resonance dominates, and the value reported is the full width of a Breit-Wigner
fit to the data.  The width for \jpsi  and \psitwos\ decaying to \epem\ 
is greater than that for \mumu\ due to the energy lost through bremsstrahlung.
\label{tab:masses}
}
\begin{center}
\begin{tabular}{ccc}
\hline\hline
Quantity   & Decay mode  & Width (\mevcc) \\ \hline
\jpsi\ mass &   \epem    &  $17 \pm 2$  \\
            &    \mumu   &  $13 \pm 1$ \\
\psitwos\ mass & \epem   &  $29 \pm 6$   \\
               & \mumu   &  $21 \pm 3$  \\
$\Delta m (\psitwos-\jpsi)$	 & \psitwos\to\jpsi\pipi ; & $7 \pm 1$  \\
                     & \jpsi\to\ellell         &   \\
$\Delta m (\chicone-\jpsi)$  & \jpsi\to\ellell         & $14 \pm 1$  \\
\KS\ mass            &   \pipi                 &  $3.5 \pm 0.2$\\
                     &   \piz\piz              &  $15 \pm 2$\\
\Kstarz\ mass        &   \Kp\pim\ and          & $60 \pm 7$ \\
                     &   \KS\piz\              &  \\
\Kstarp\ mass        &   \KS\pip\ and          & $50 \pm 10$ \\
                     &    \Kp\piz              &    \\ \hline\hline
 \end{tabular}
\end{center}
\end{table}

The next step in the analysis is to combine sets of tracks and/or neutral 
clusters to form candidates for the initial or intermediate mesons in
the decay.  Our general strategy when forming these candidates
 is to assign the expected masses to tracks and neutral clusters, and to apply
a vertex constraint before computing the invariant mass.  
In rare instances (less than 1\% of all meson candidates) the vertex fit does not 
converge.  The sum of the track and/or cluster four-vectors
is used to compute the invariant mass for such candidates.
If one or more decay products from a given particle are themselves 
intermediate states, we 
constrain them to their known masses.
At each step in the decay chain, we require that mesons have
masses consistent with their assumed particle type.
The mass resolutions observed for all of the intermediate 
mesons considered in this 
paper are listed
in Table~\ref{tab:masses}.

We choose meson selection criteria to 
maximize the expected precision of our branching fraction measurements.  
Therefore we use well-understood quantities in our selection, 
which lead to a 
smaller systematic
uncertainty.  We set the selection values to maximize the 
ratio $S/\sqrt{S+B}$ where 
$S$ and $B$ are the expected number of signal and background events 
respectively, 
as estimated from Monte Carlo.  If a given mode has been 
previously observed, $S$ is estimated using the known 
branching fraction.
Otherwise, selection values 
similar to those 
in previously-observed modes are taken as a starting point, and then
modified to reduce
background (as measured in the kinematic sidebands) 
or increase signal efficiency (as measured using Monte Carlo simulated signal 
events).
In most cases, we find that $S/\sqrt{S+B}$ does not change 
significantly when selection values are varied near their optima.  
This allows
us to choose standard selection values across most final states.

%
\subsection {Charmonium Meson Candidate Selection}

\begin{figure}
\begin {center}
\epsfig{file=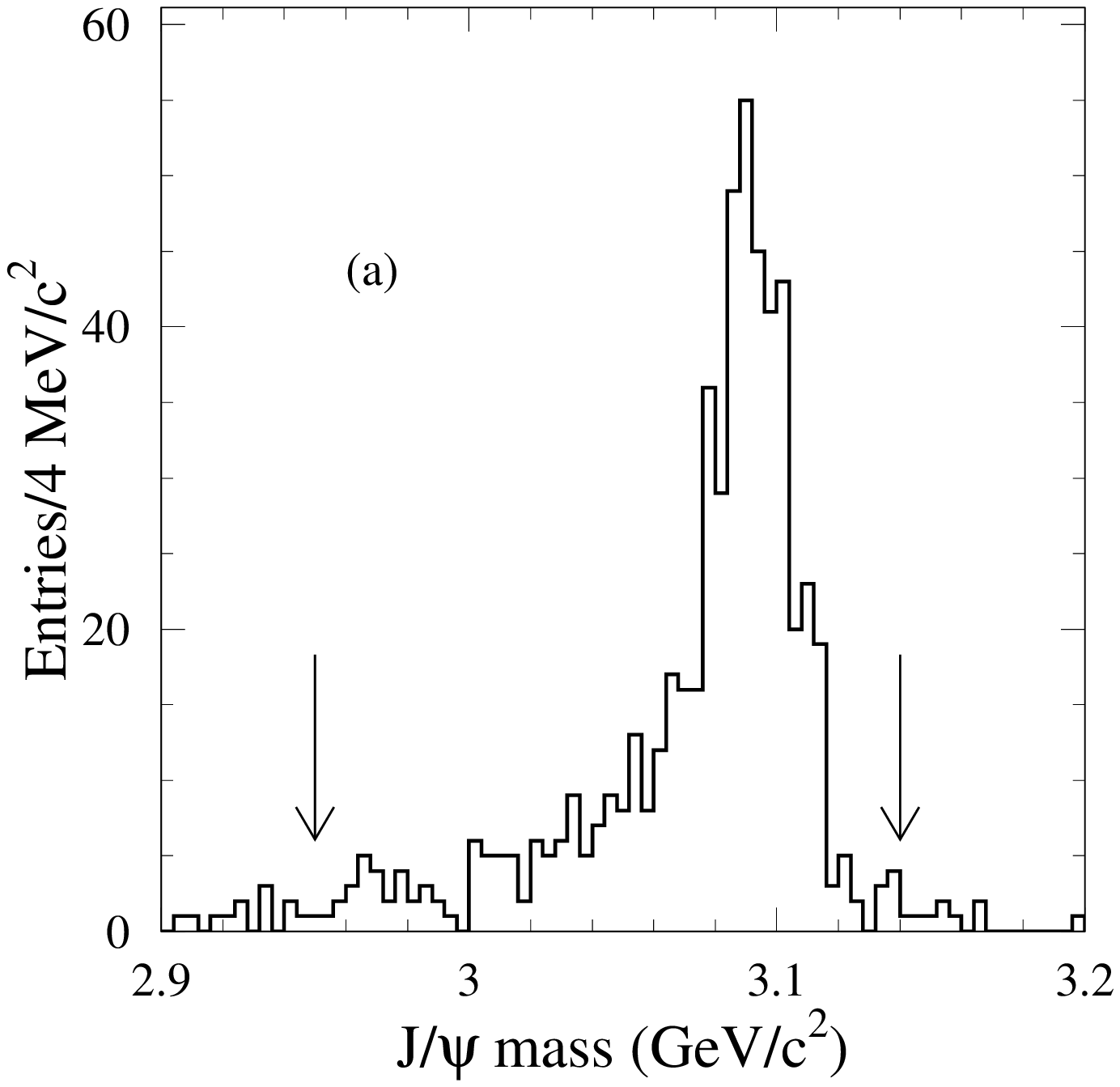, width = 0.4\textwidth} \\
\epsfig{file=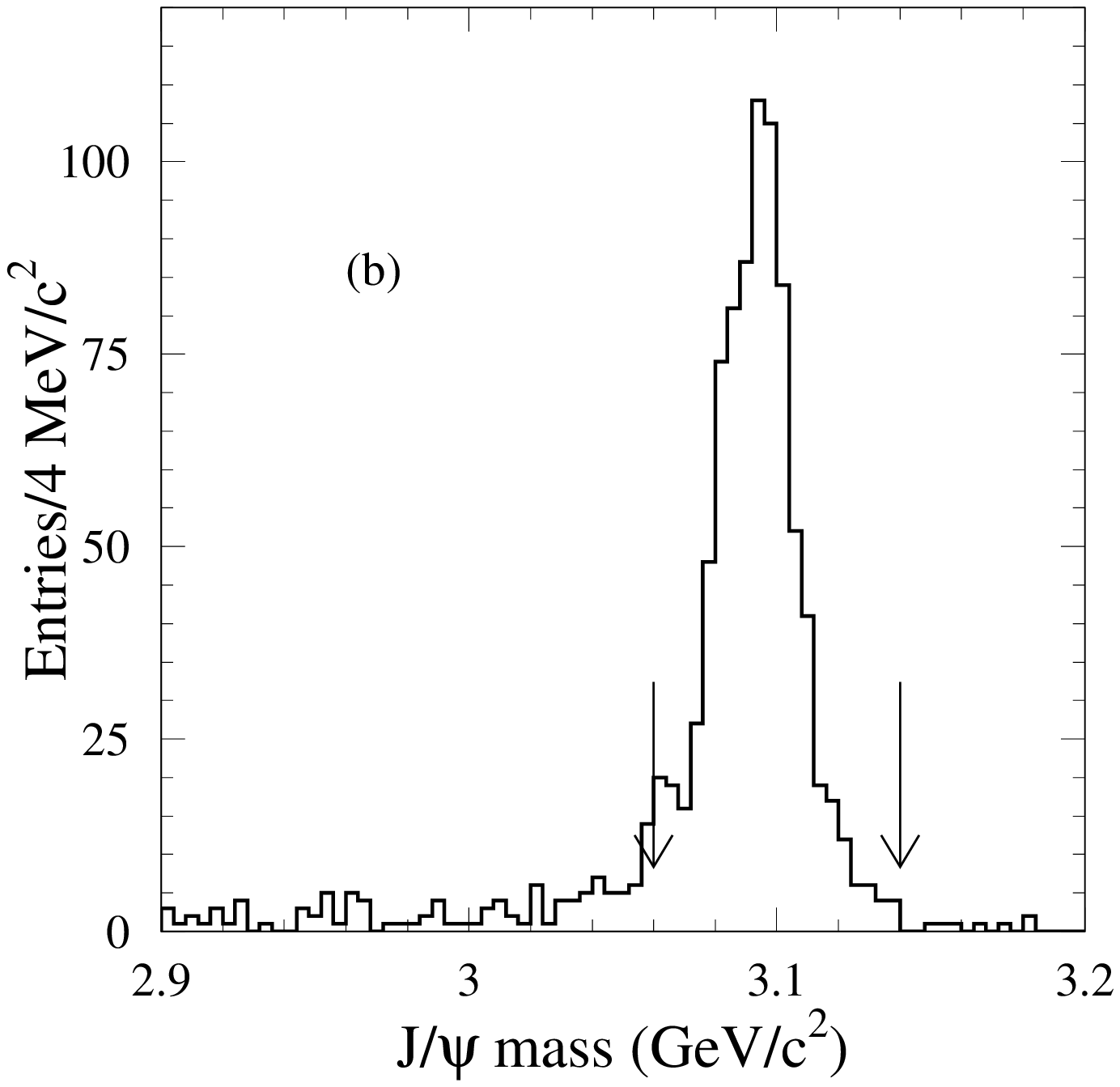, width = 0.4\textwidth} \\
\caption{Invariant mass distribution for (a) \jpsi\to\epem and 
(b) \jpsi\to\mumu\  candidates mass 
observed in \bzjpsiks\ and \bpjpsikp\ candidates passing
the exclusive branching fraction selection.   The mass interval used
to select \jpsi\ candidates for $B$ reconstruction is
indicated by the arrows.
\label{fig:jpsimass}}
\end{center}
\end{figure}

\begin{figure}
\begin {center}
\epsfig{file=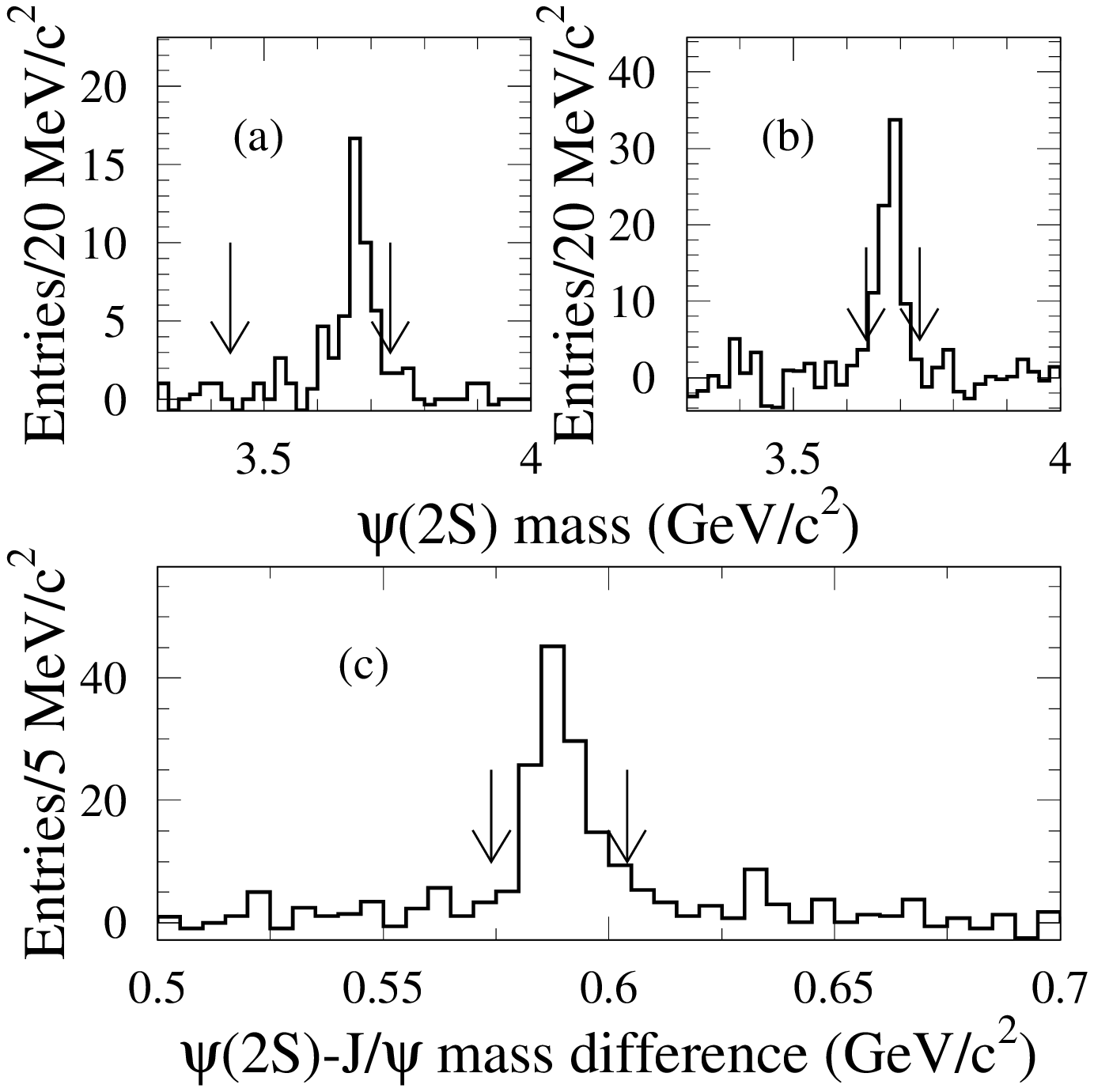, width = 0.4\textwidth}
\caption{Background-subtracted \psitwos\ candidate mass and mass 
difference distributions 
observed in \bzpsitwosks\ and \bppsitwoskp\ candidates passing
the exclusive branching fraction selection, for (a) \psitwos\to\epem, 
(b) \psitwos\to\mumu, and (c) the \psitwos-\jpsi\ mass difference 
distribution for
\psitwos\to\jpsi\pipi\.   The intervals used
to select \psitwos\ candidates for $B$ reconstruction are
indicated by the arrows.
\label{fig:psitwosmass}}
\end{center}
\end{figure}

\begin{figure}
\begin {center}
\epsfig{file=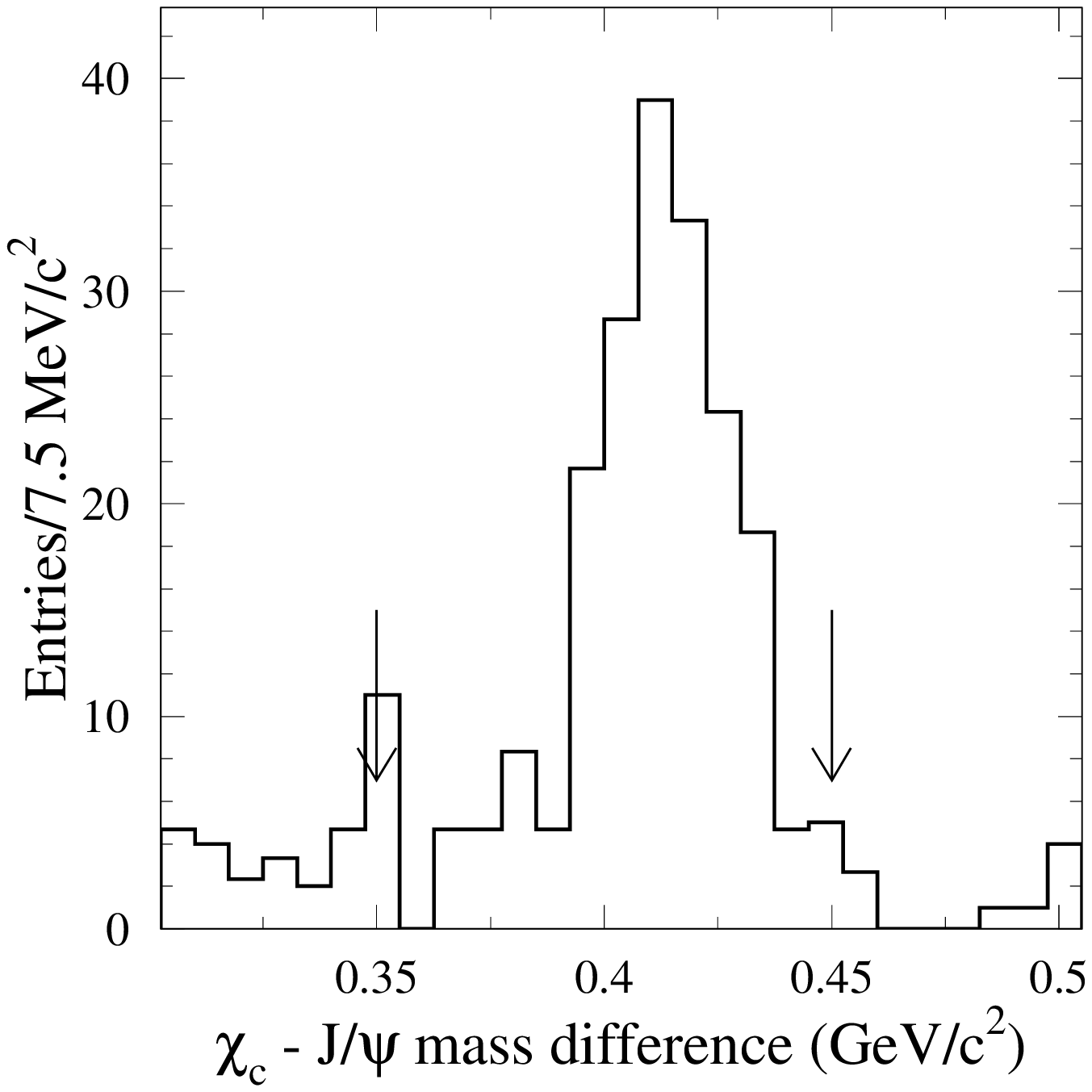, width = 0.4\textwidth}
\caption{Background-subtracted \chic{1}-\jpsi\ candidate mass 
difference distribution
observed in \bzchiconeks\ and \bpchiconekp\ candidates passing
the exclusive branching fraction selection.  The mass difference interval used
to select \chic{1}\ candidates for $B$ reconstruction is
indicated by the arrows.
\label{fig:chicmass}}
\end{center}
\end{figure}

\subsubsection{\jpsi\ Selection}

\label{sec:jpsi}
 \jpsi\ candidates are required to have an invariant mass in the range
$2.95 <M_{\jpsi} < 3.14 $\gevcc  and $3.06 < M_{\jpsi} < 3.14 $\gevcc for
\jpsi\to\epem and  \jpsi\to\mumu decays respectively.
Unless otherwise stated, for \jpsi\to\epem decays, one track is
required to pass the tight electron selection and the other the loose
selection.  Tracks not associated to an EMC cluster that pass the DCH-only
selection are also accepted. For 
\jpsi\to\mumu decays, we require one track to pass the loose selection and
the other to pass the MIP selection.

The mass distribution for \jpsi\ candidates in the data is 
shown in Fig.~\ref{fig:jpsimass}.

%

\subsubsection{\psitwos\ Selection}
\label{sec:psitwos}

\psitwos\to\mumu\ candidates are required to have
a mass within 50 \mevcc\ of the known
 \psitwos\ value of 3.69~\gevcc \cite{PDG2000}. 
For \psitwos\to\epem\ candidates
the lower bound is relaxed to 250 \mevcc below the 
known value.  
For decays of the \psitwos\ to \jpsi\pipi, the difference in 
mass between the \psitwos\ and \jpsi\ candidates is required to be
within 15 \mevcc\ of the expected value, and the \pipi\ invariant 
mass $m_{\pipi}$ is required to be between 0.4 and 0.6 \gevcc.  
The latter 
requirement takes advantage of the fact that $m_{\pipi}$ is most often
in the upper portion of the kinematically allowed range~\cite{BESpsi2S}.   
All \psitwos\ candidates are required to have a momentum in the 
center-of-mass frame
between 1.0 and 1.6 \gevc, consistent with $B \to \psitwos K$ decays.  

We have used the same lepton identification requirements
as for the \jpsi 
reconstruction. These are applied either to the leptons from  
\psitwos\to\ellell\ decays, or to the leptons from the \jpsi\ in 
\psitwos\to\jpsi\pipi\ decays.

The mass and mass difference distributions for \psitwos\ candidates 
in the data are 
shown in Fig.~\ref{fig:psitwosmass}.  For Figures~\ref{fig:psitwosmass},
\ref{fig:chicmass}, and  \ref{fig:kstarmass} a background subtraction is
performed using the observed distribution of candidates in the \De\ sidebands
(see Section~\ref{sec:ExclB}).

%
%
\subsubsection{\chic{1} \ Selection}
\label{sec:chicone}

In reconstructing $\chic{1}\to\jpsi\gamma$, 
\jpsi\ and photon candidates are
selected as described above.
The muon identification requirements are subsequently tightened 
by demanding that one lepton from the \jpsi\ pass 
the loose selection and the other the very loose selection 
(rather than the MIP selection).

In addition, the photon cluster is required to satisfy
$E > 150 {\rm\mev}$ and $A_{42}  < 0.15$ 
and to have a centroid in the angular range $0.41  <  \theta  < 2.409$, 
excluding the forward direction due to the
increased material (from electronics, cables, and final-focusing magnets) in
that region. 

We require the mass difference between the reconstructed 
\chic{1}\ and \jpsi\ candidates to satisfy
$0.35  <  M_{ \g\jpsi }-M_{\jpsi}  <  0.45 \ {\rm\gevcc}$.

The mass difference distribution for \chic{1} candidates in the data is 
shown in Fig.~\ref{fig:chicmass}.
%
%
%
\subsection{Light Meson Candidate Selection}

\begin{figure}
\begin {center}
\epsfig{file=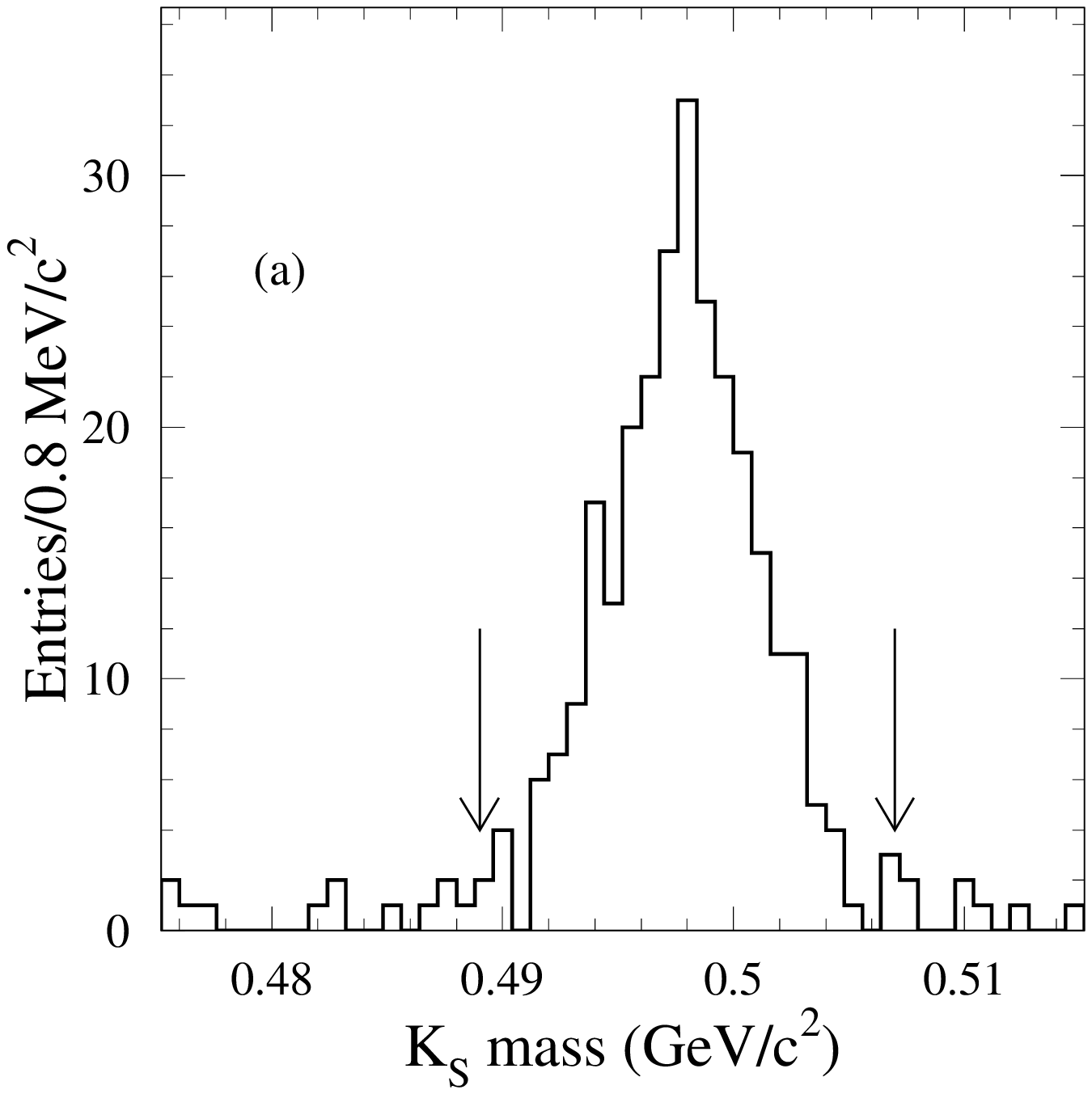, width = 0.4\textwidth} \\
\epsfig{file=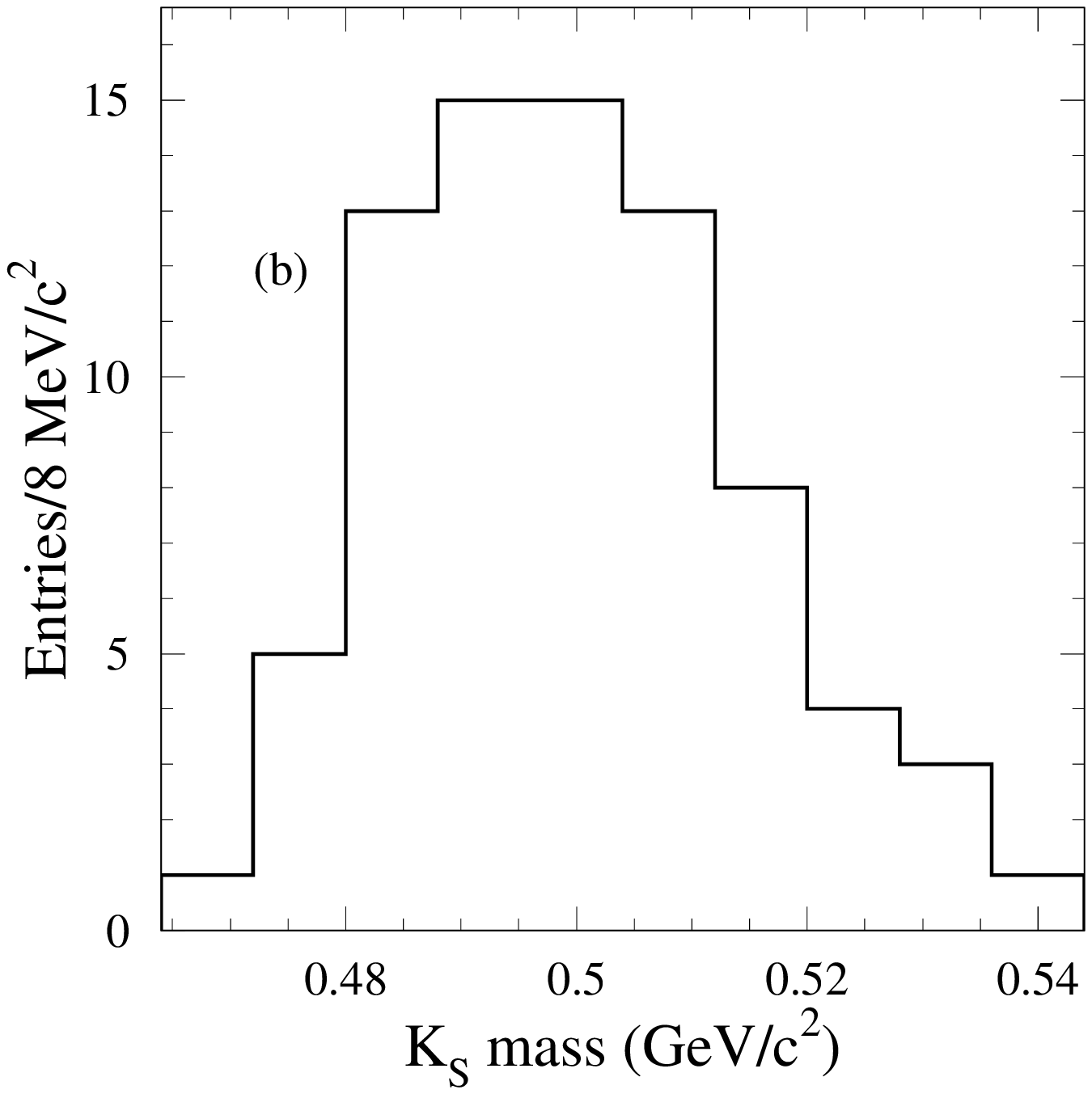, width = 0.4\textwidth} \\

\caption{\KS\ candidate mass distribution observed in \bzjpsiks\ candidates
 passing
the exclusive branching fraction selection, for (a) \KS\to\pipi\  and
(b) \KS\to\ppz.   The mass intervals used
to select \KS\to\pipi\ candidates for $B$ reconstruction is
indicated by the arrows in (a); the full range of (b) is used in selecting
\KS\to\ppz\ candidates.
\label{fig:ksmass}}
\end{center}
\end{figure}

\begin{figure}
\begin {center}
\epsfig{file=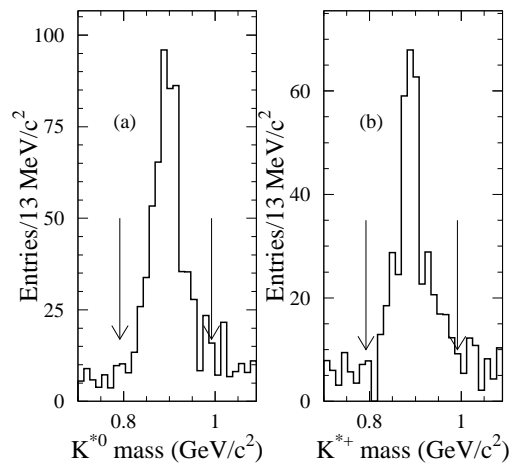, width = 0.40\textwidth} 

\caption{Background-subtracted (a) \Kstarz\ and (b) \Kstarp\ candidate mass 
distributions observed in \bzjpsikstarz\ and \bpjpsikstarp\ candidates passing
the exclusive branching fraction selection.  The mass interval used
to select \Kstar\ candidates for $B$ reconstruction is
indicated by the arrows.
\label{fig:kstarmass}}
\end{center}
\end{figure}

\subsubsection{\piz\to\gaga\ Selection}
We reconstruct \piz\ candidates as pairs of photons.  Individual photons
separated by distances of 10 cm or
more in the EMC are reconstructed as distinct clusters.  
Photons from \piz's with energies above 2 \gev\ can have 
less separation, in which case the two photons 
are
reconstructed as a single cluster.  We refer to these as ``merged'' \piz's.
 They are distinguished from single photons based on their shower shape.

\subsubsection{\KS\to\pipi\ Selection}
\label{sec:kscharged}
We construct \KS candidates from all pairs of oppositely 
charged tracks, and retain those that have invariant mass
between 489 and  507 \mevcc\ after applying a vertex constraint.    
To further reject background 
we exploit the flight length of the \KS\ by demanding that the \KS\
 vertex be more than 1 \mm\ (in three dimensions) 
from the \jpsi, \psitwos, or \chic{1}\ vertex.

The mass distribution for \KS\to\pipi\ candidates in the data is 
shown in Fig.~\ref{fig:ksmass}.


%
%
%
\subsubsection{\KS\to\piz\piz\ Selection}
\label{sec:ksneutral}

The $\KS \rightarrow \ppz \rightarrow 4\gamma$ decay chain is
reconstructed
from photon combinations satisfying
$E_{\rm \gamma}>30 \mev$, $E_{\piz}>200 \mev$ and
$E_{\KS}>800\mev$,
with $110 \leq m_{\piz} \leq 155 \mevcc$ and $300 \leq m_{\KS} \leq 800 \mevcc$.
We perform a mass-constrained fit to each photon pair with the known 
\piz\ mass.  This fit is repeated assuming different decay
points
along the \KS\ flight path, as defined by the \jpsi\ vertex and the
initial \KS\
momentum vector direction.
The point where the product of the fit $\chi^2$ probabilities
for the two \piz 's is maximal is defined as the \KS\ decay vertex.
\KS\ candidates with flight length in the range from $-10$ to +40 cm
 are retained.

We consider merged \piz\ candidates with energy above 1\gev.
If an EMC cluster candidate is identified as a merged \piz\ 
but can also be paired with another photon to form a \piz\ candidate, 
we use the latter interpretation.  Merged \piz's represent less than 10\% of
all \piz's used in this analysis.

The invariant mass of the \KS\ candidate at the optimal vertex point is required 
to lie in the range 470 to 550\mevcc .

The mass distribution for \KS\to\ppz candidates in the data is 
shown in Fig.~\ref{fig:ksmass}.

%
\subsubsection{\Kstarz\ and \Kstarp\ Reconstruction}
\label{sec:kstar}
We reconstruct the \Kstarz\ through its decays to $\Kp\pim$ and 
$\KS\piz$ and the \Kstarp\ through its decays to $\KS\pip$ and $\Kp\piz$,
where the \KS\ is reconstructed in the \pipi\ mode.

\piz's are reconstructed from isolated photons and 
required to have an invariant mass between 106 and 153\mevcc.
If there is a \KS\ in the final state
we require that the angle in the $xy$ plane between the
\KS\ momentum vector and the line joining 
the \jpsi\ and \KS\ vertices be less than
200\mrad\ and that 
 the \KS\ 
vertex fit converge.

In addition, for channels containing a \piz\ in the final state, we demand that the 
cosine of the angle $\theta_K$, measured in the \Kstar\ rest frame, 
between the kaon 
momentum and the \Kstar\ direction as measured in the $B$ frame be less 
than 0.95.  


All candidate \Kstar's\ are required to be within 100\mevcc\ of the known
\Kstarz\ or \Kstarp\ mass~\cite{PDG2000}.

The mass distribution for \Kstar\ candidates in the data is 
shown in Fig.~\ref{fig:kstarmass}.

%
%
%

\subsection{$B$ Meson Candidate Selection}
\label{sec:ExclB}

\label{sec:helicity}
\begin{figure}
\begin{center}   
\epsfig{file=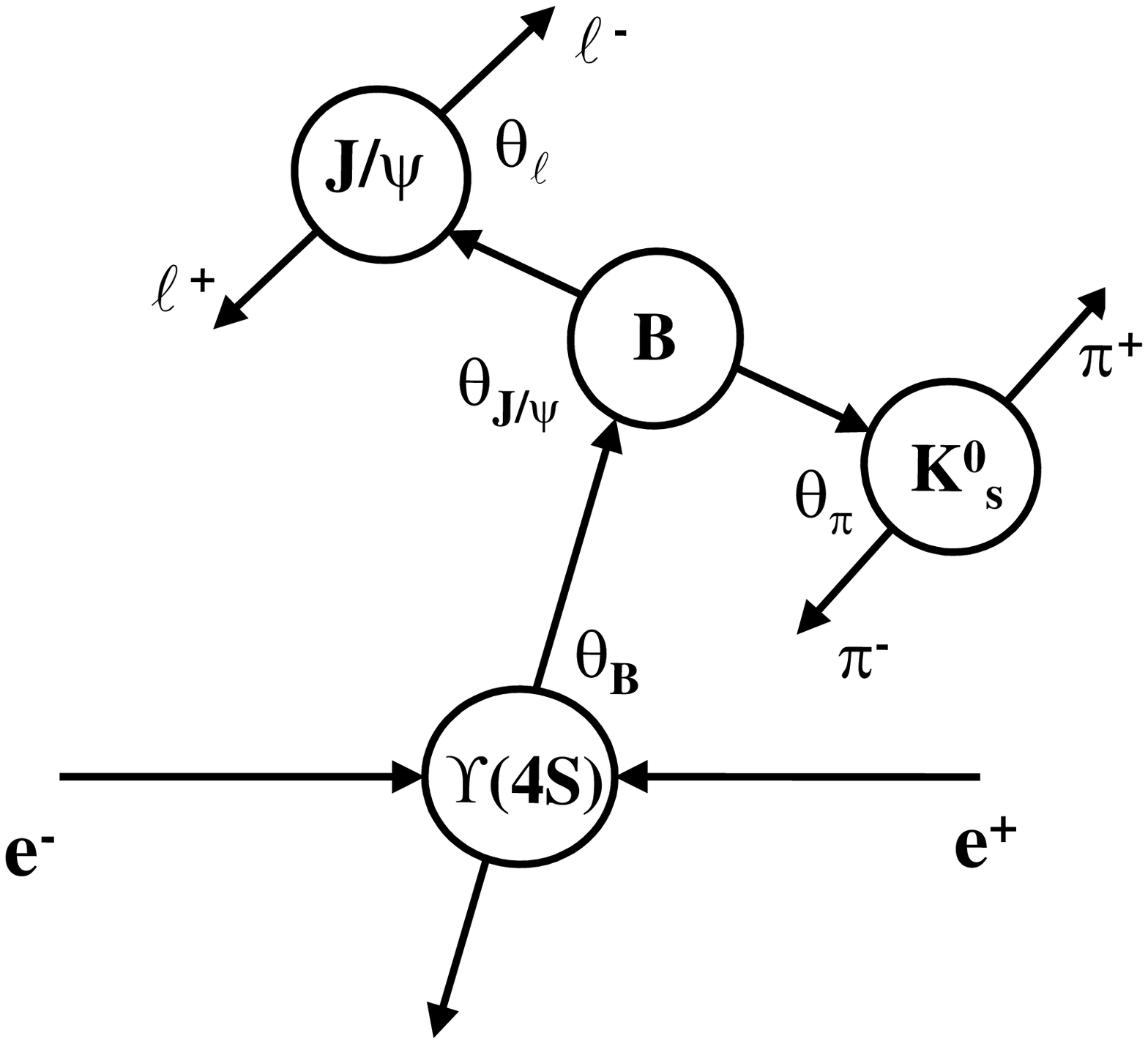, width=0.45\textwidth} 
\end{center}
\caption[Helicity]{Helicity angles for the decay \FourS\to\BB\to\jpsi(\epem or \mumu)+\KS.
\label{fig:helicity}}
\end{figure}

\begin{figure}
\begin{center}   
\epsfig{file=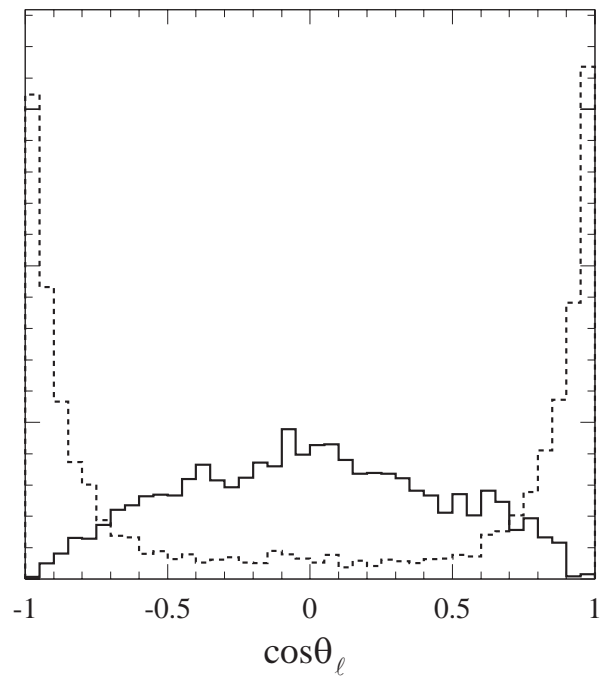, width=0.45\textwidth} 
\end{center}
\caption[cosheli]{Distributions of $\cos\theta_\ell$ observed in \bzjpsiks\
and \bpjpsikp\ candidates.  The dashed histogram shows candidates
in the \De\ sideband.  The solid histogram shows the distribution in the
\De-\mes\ signal region, after subtracting the distribution observed in the
sideband scaled by the ratio of signal to sideband areas. 
The normalization of both histograms has been set to unity.
\label{fig:cosheli}}
\end{figure}

\begin{table*}
\caption{\label{tab:sigbox}
Definition of the signal region in $|\De|$ and \mes\ for each mode used in
this analysis. The  \mes signal region is given in terms of $|\mes - m_B|$, 
where $m_B$ is 5279 \mevcc.}
\begin{center}
\begin{tabular}{ccccc}
\hline\hline
$B$ decay    & Light meson & Charmonium meson & $|\De|$ (\mev) & $|\mes - m_B|$  (\mevcc) \\
mode         & decay mode  & decay mode       &              &      \\ \hline
\bzjpsiks    & \pipi       & \epem            &  34.5        &     8.1     \\
             &             & \mumu            &  29.0        &     7.2     \\
             & \ppz        & \epem            & 100.0        &     8.0     \\
             &             & \mumu            & 100.0        &    10.0     \\
\bzjpsikl    &   --        & \epem \& \mumu   &  10.0        &     --      \\
\bpjpsikp    &   --        & \epem            &  38.4        &     7.5     \\
             &             & \mumu            &  30.3        &     6.9     \\
\bzjpsikstarz& \Kp\pim     & \epem            &  30.9        &     9.3     \\
             &             & \mumu            &  23.7        &     8.1     \\
             & \KS\piz     & \epem            &  48.6        &    12.0     \\
             &             & \mumu            &  45.6        &    11.4     \\
\bpjpsikstarp& \KS\pim     & \epem            &  62.7        &     7.2     \\
             &             & \mumu            &  20.4        &     9.9     \\
             & \Kp\piz     & \epem            &  85.2        &    11.4     \\
             &             & \mumu            &  50.1        &    10.2     \\
\bzjpsipiz   & \gaga       & \epem \& \mumu   &  112.0       &     9.0     \\
\bzpsitwosks & \pipi       & \epem \& \epem\pipi & 28.0      &     9.0     \\
             &             & \mumu \& \mumu\pipi & 26.0      &     9.0     \\
\bppsitwoskp & \pipi       & \epem \& \epem\pipi & 28.0      &     9.0     \\
             &             & \mumu \& \mumu\pipi & 26.0      &     9.0     \\
\bzchiconeks & \pipi       & \epem$\gamma$    & 30.9         &     6.9     \\
             &             & \mumu$\gamma$    & 21.4         &     6.9     \\
\bpchiconekp & \pipi       & \epem$\gamma$    & 33.9         &    11.7     \\
             &             & \mumu$\gamma$    & 27.9         &     6.6     \\
\bzchiconekstarz & \Kp\pim & \epem$\gamma$    & 30.0         &     9.0     \\
             &             & \mumu$\gamma$    & 30.0         &     9.0     \\
\hline\hline
\end{tabular}
\end{center}
\end{table*}

$B$ mesons are reconstructed by combining charmonium meson 
candidates with light meson candidates.  Both the charmonium and light meson
candidates are constrained to their known masses, with the exception
of \Kstar\ candidates, for which the natural width dominates the experimental
resolution.   
Two kinematic variables are used to isolate the $B$ meson signal for all
 modes except \bzjpsikl.
One is the difference between the reconstructed energy of 
the $B$ candidate and the beam energy in the center-of-mass frame \De.
  The other is the beam energy substituted mass \mes, defined as:
\begin{equation}
  \mes =  \sqrt{E_{\rm beam}^{*2}-p_B^{*2}}
\end{equation}
where $p_B^*$ is the momentum of the reconstructed $B$ and $E_{\rm beam}^*$ 
is the beam energy, both in the center-of-mass frame.
The small variations of $E_{\rm beam}^{*}$ over the duration of the run 
are taken into account when calculating \mes.  Signal events will have 
\De\ close to 0 and \mes\ close
to the $B$ meson mass, 5.279 \gevcc.

We limit all our two dimensional plots in these variables to the ``signal neighborhood'',
defined by $\mid$\De$\mid<\Delta E_{\rm max}$ and $5.2<\mes<5.3$ \gevcc.
For most channels, $\Delta E_{\rm max}$ is 120 \mev, but for the
\bzjpsiks (\KS\to\piz\piz) and \bzjpsipiz\  channels, which have
larger \De\ resolution, it is increased to 150 and 400 
\mev\ respectively. 
  We define the 
signal region by fitting the observed distribution of events in the signal 
neighborhood in \mes\ and \De\ separately.  In the fit, the signal
component is modelled by a Gaussian, and the background component is modelled
by an empirical phase-space 
distribution ~\cite{ARGUS_bkgd} (henceforth referred to as the ARGUS
distribution) when fitting the \mes\ distribution,  or a polynomial
when fitting the \De\ distribution.  The ARGUS distribution is 
\begin{eqnarray}
\label{eq:ARGUS}
A(\mes;m_0,c)& \propto &\mes\sqrt{1-(\mes/m_0)^2)} \times \nonumber \\
             &         & \exp(c(1-(\mes/m_0)^2)),
\end{eqnarray} 
where $m_0$ is set to a typical beam energy and $c$ is a fitted parameter.

The widths of the fitted Gaussians provide a
measurement of the resolution in \De\ and \mes, and the signal region is 
defined as $\pm 3\sigma$ about the nominal value in each variable.
The resolution in \mes\ is typically 3 \mevcc, and that in \De\ is typically
10 \mev\ for channels with no neutral particles in the final state and 30 \mev\ otherwise.  The signal region for each mode is given in 
Table~\ref{tab:sigbox}.

A somewhat different procedure is required for reconstructing \bzjpsikl, 
since the \KL\ energy is not measured.  Either the $B$ mass
 or energy must be constrained, leaving only one 
independent variable.  We choose to fix the $B$ mass to its known
 value~\cite{PDG2000} and plot the
signal in the quantity \DeKL $\equiv E_{\jpsi}^* + E_{\KL}^* - E^*_{\rm beam}$, where 
$E_{\jpsi}^*$ is the energy of the mass-constrained \jpsi, 
and $E_{\KL}^*$ is the energy of the  \KL\
as determined using the $B$ mass constraint, both in the center-of-mass 
frame.  \DeKL\ is a measure of the same
quantity as \De; we use the different notation to reflect the fact that the
$B$ mass constraint is used only in this channel. 
  For signal, \DeKL\ is expected 
to peak at zero with a resolution of approximately 3.5~MeV.
The signal region is defined 
as $|\DeKL | < 10 \mev$.  

\subsubsection{Helicity and Thrust Angle Definitions}
We use the 
helicity angles $\theta _{B}$ and $\theta _{\ell }$ to help distinguish 
signal from background.  
$\theta _{B}$ is the angle in the center-of-mass frame between the electron 
beam and $B$ candidate directions, and $\theta _{\ell }$ is the angle in the
charmonium meson rest frame between the $\ell^-$ and light meson candidate directions.  
Figure~\ref{fig:helicity} gives a 
schematic representation of  these 
angles for 
the decay \bzjpsiks.

The angle $\theta _{B}$ has a $\sin^2\theta_B$ distribution for 
\FourS\ meson decays.  If $X$ is a pseudoscalar (\Kz, \Kp, \piz) 
then the charmonium meson must 
be longitudinally polarized, and the resulting $\theta _{\ell}$
distribution is proportional to $\sin ^{2}\theta _{\ell}$. 
If $X$ is a vector ($\Kstar$) 
the decay angular distribution depends on more than one helicity amplitude. 
In this case the
lepton angular distributions are not known {\em a priori} and must be 
experimentally
determined. 

The $B$ candidates formed from light
quark backgrounds will generally follow a $1+\cos ^{2}\theta _{B}$ angular
distribution. The  $\theta _{\ell}$ helicity angle is especially useful in
rejecting background since the distribution of $\cos \theta _{\ell}$ is
peaked at $\pm 1$ for background and at zero for signal for modes where $X$
is a pseudoscalar.  As an example, the distribution of $\cos \theta _{\ell}$
observed in data for \bzjpsiks\ and \bpjpsikp\ candidates is shown in 
Fig.~\ref{fig:cosheli}.

For modes where the charmonium meson decays to more than two bodies, and
 $\theta_\ell$ is therefore undefined, we suppress backgrounds using the 
thrust angle $\theta_T$, defined as the angle between the thrust axis
of the 
reconstructed \B\ and
that of the rest of the event in the center-of-mass frame.   
We use the conventional definition of the thrust axis for a collection 
of particles
 as the direction
about which the transverse momenta of the particles is minimized.
In \BB\ events $\cos\theta_T$ is uniformly distributed, whereas in 
continuum background 
events $\theta_T$ tends
to peak at $\pi$ radians due to the two-jet nature of
these events.  Hence $\theta_T$ can be used to discriminate against background in modes
where the helicity angle is not applicable.

 The helicity and thrust angle values used to select candidates are listed 
in the appropriate exclusive
reconstruction and selection subsections in this paper.

\subsubsection{Multiple Candidates}
We only allow one exclusive candidate per event in a 
given decay mode.
In the cases where we have multiple candidates (less than 10\% of all events
with a candidate for most modes, but up to 30\% for the \Kstar\ modes which have
significant crossfeed among decay channels), 
the candidate with the lowest $\mid$\De$\mid$ is taken over all others. 
The only exception is in the \bzjpsikl\ selection, where
we choose the candidate with the largest \KL\ energy as
measured by the EMC.  If none of the candidate  \KL\ mesons have EMC 
information, we choose 
the candidate that has 
the largest number of layers with hits in the IFR.
These criteria are used because background candidates tend to have 
low EMC energy or IFR multiplicity. 
%
%
\begin{figure}
    \begin{center}
       \epsfig{file=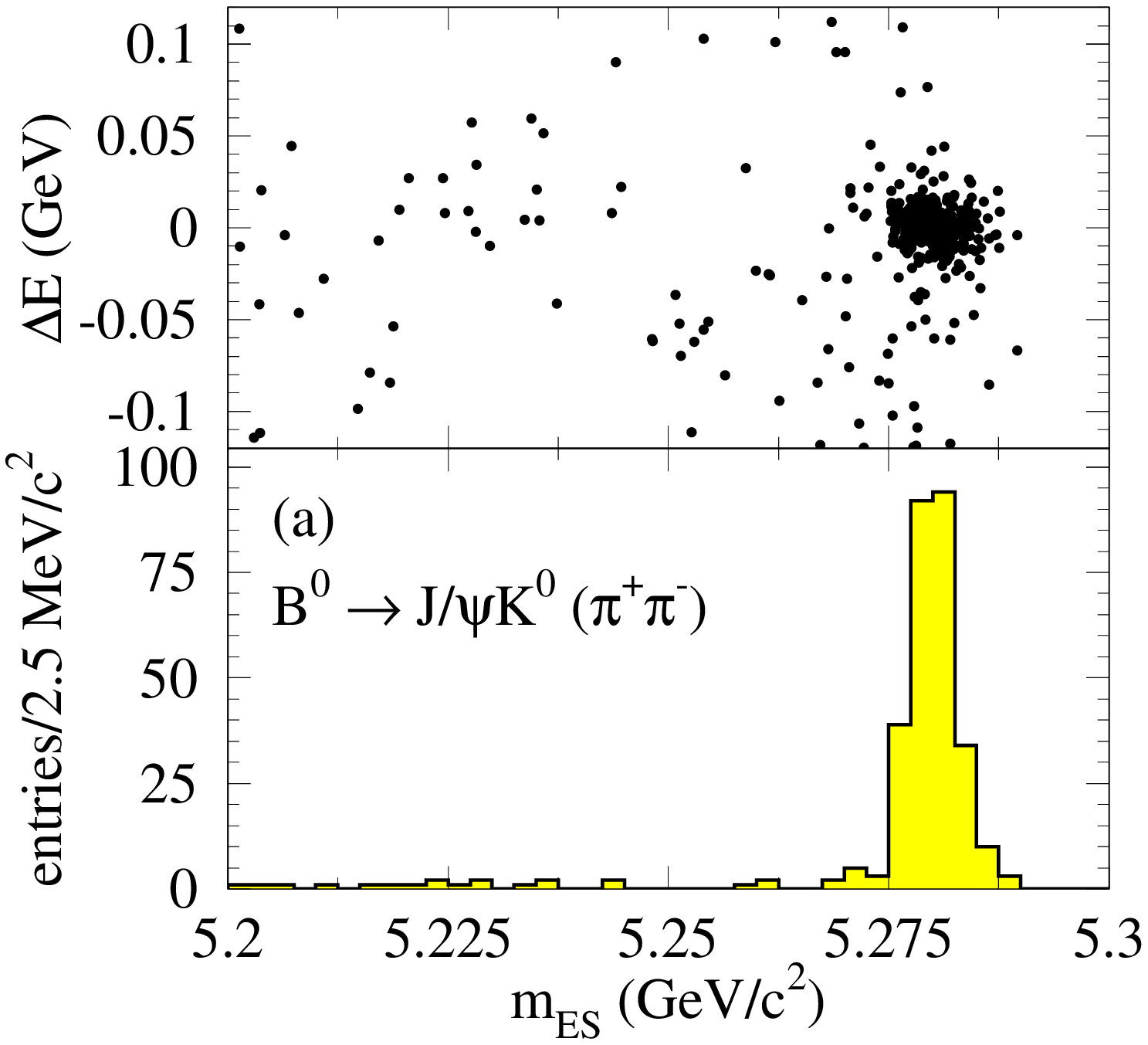,width=0.35\textwidth} \\ \vspace{-0.5cm}
       \epsfig{file=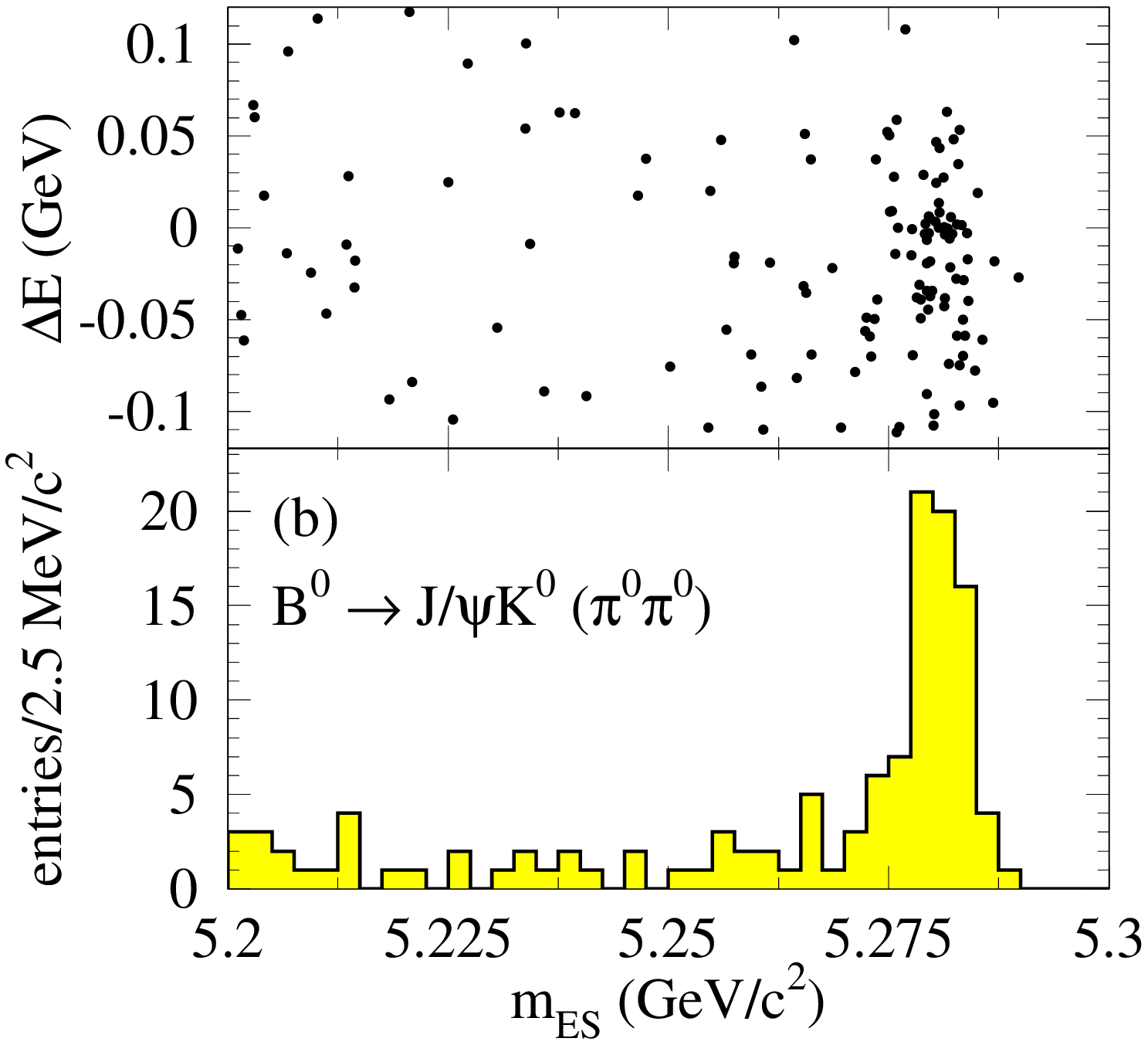,width=0.35\textwidth} \\ \vspace{-0.5cm}
       \epsfig{file=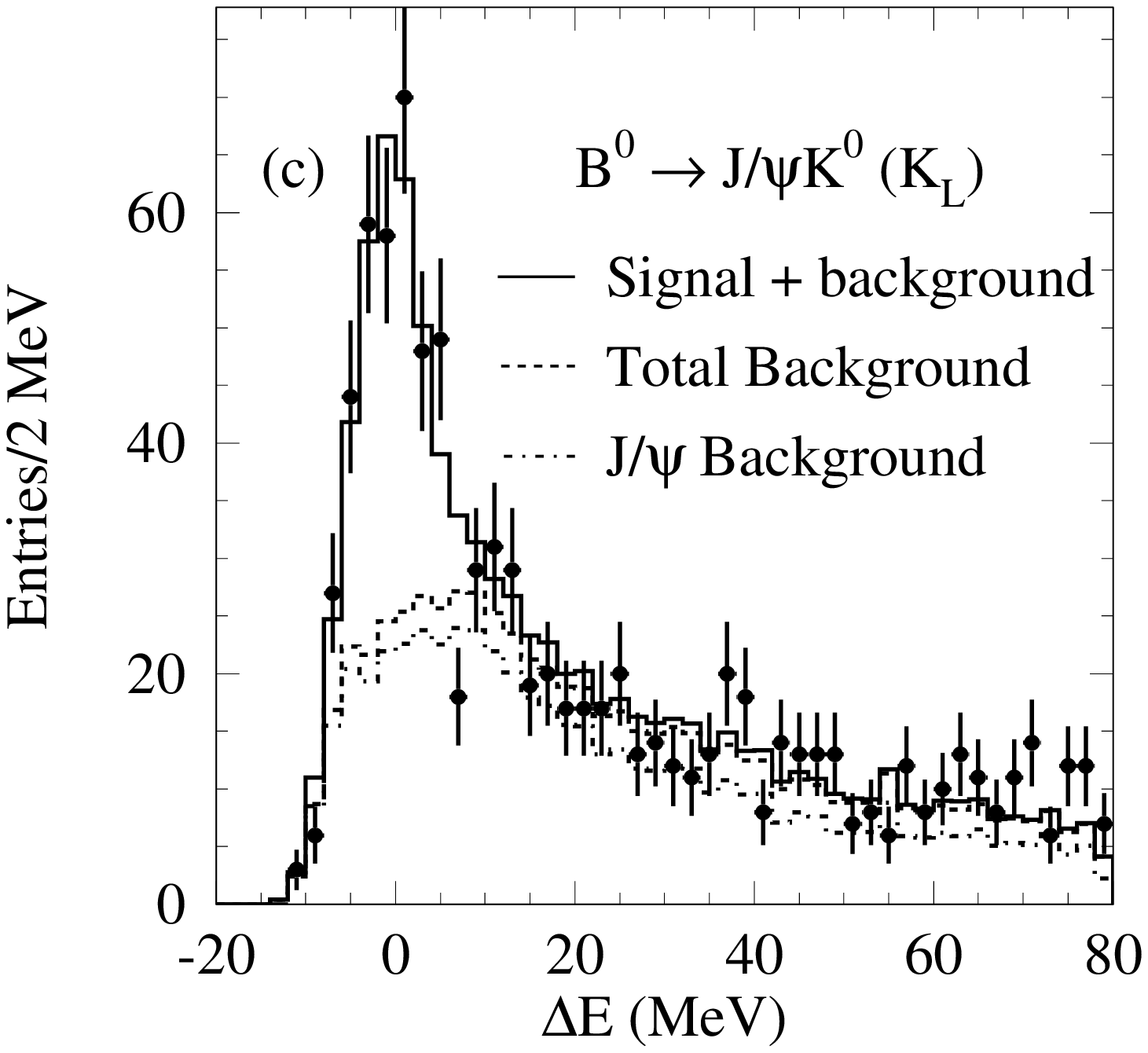,width=0.35\textwidth}   
      \caption[fig:JpsiK0]{
Signals for \bzjpsiks\ ((a) \KS\to\pipi\ and (b) \KS\to\piz\piz) and 
(c) \bzjpsikl.  In (a) and (b) the upper plots show the distribution of 
events in the \De-\mes\ plane, and the lower plots show the distribution in
\mes\ of events in the signal region in \De.  In (c) the points are the data,
the dot-dashed line shows the Monte Carlo simulated distribution of background 
events which include a real \jpsi, the dashed line shows the model for the 
total background, where
the non-\jpsi\ component is taken from the \jpsi\ sidebands in data, and the
solid line shows the sum of the background and signal Monte Carlo models.
\label{fig:JpsiK0}}
  \end{center}    
\end{figure}

\begin{figure*}
    \begin{tabular}{lcr} \hspace{-1cm} 
       \epsfig{file=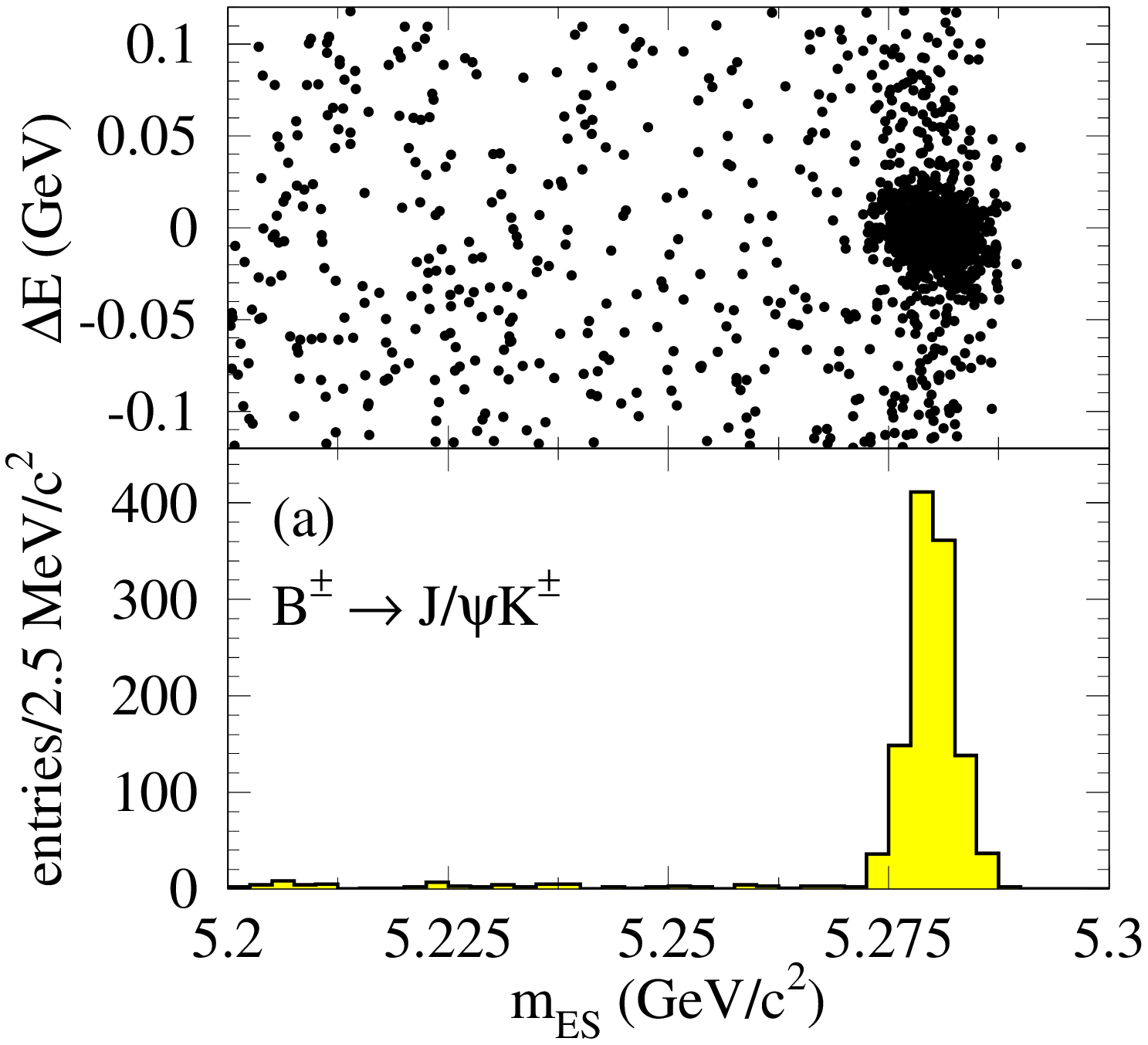,width=0.35\textwidth} &\hspace{-0.25cm} 
       \epsfig{file=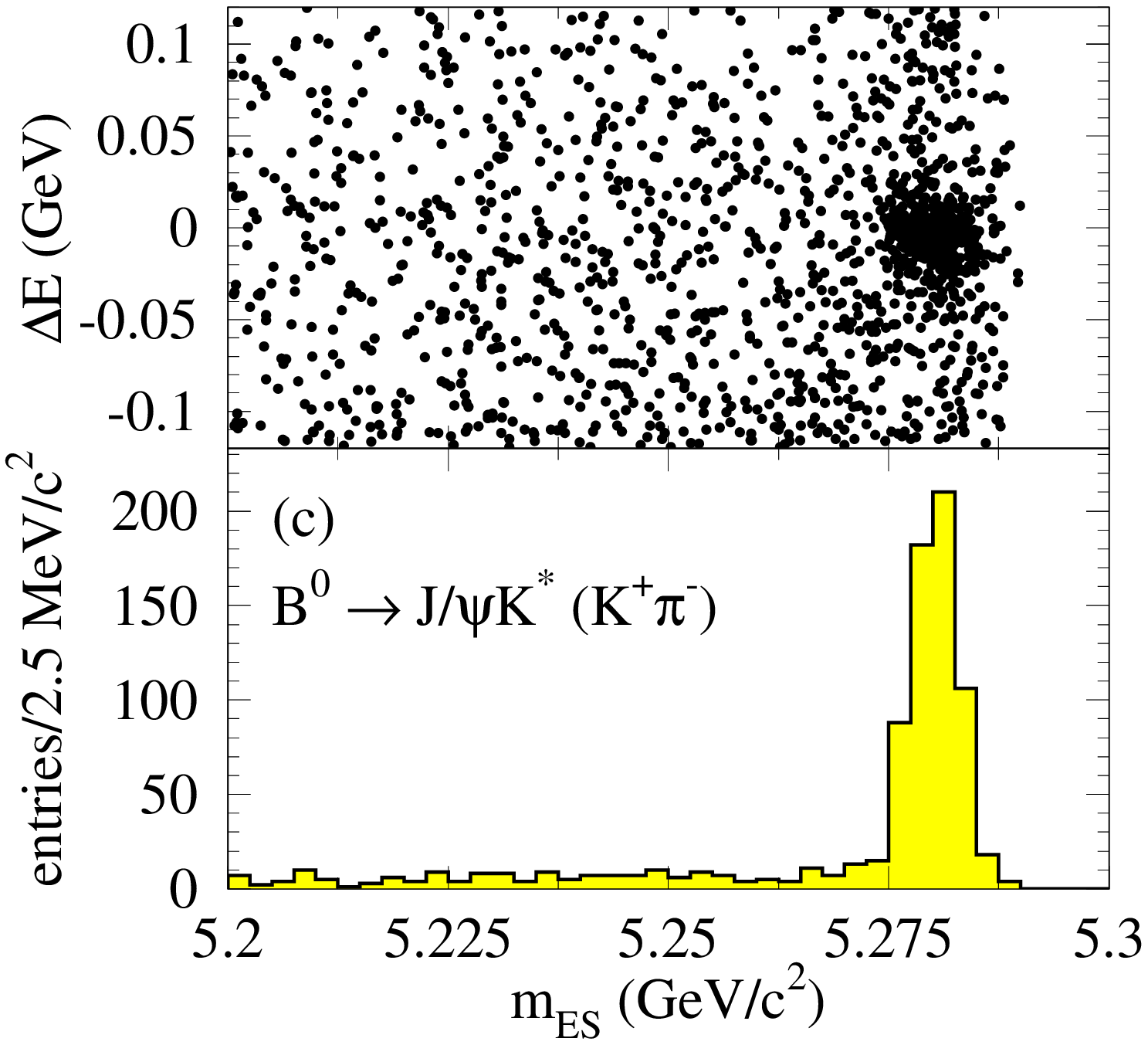,width=0.35\textwidth} &\hspace{-0.25cm} 
	\epsfig{file=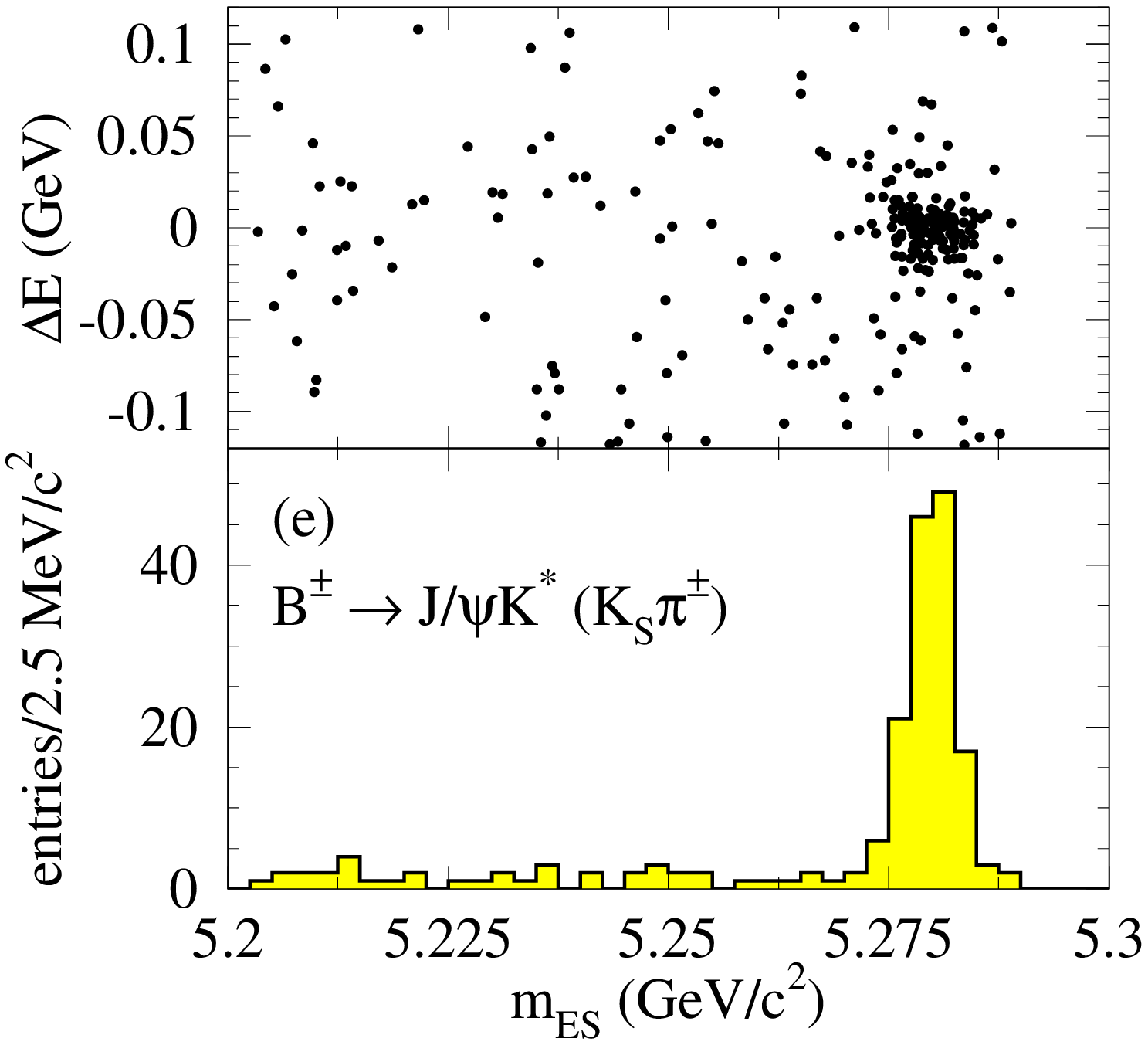,width=0.35\textwidth} \\ \hspace{-1cm} 
       \epsfig{file=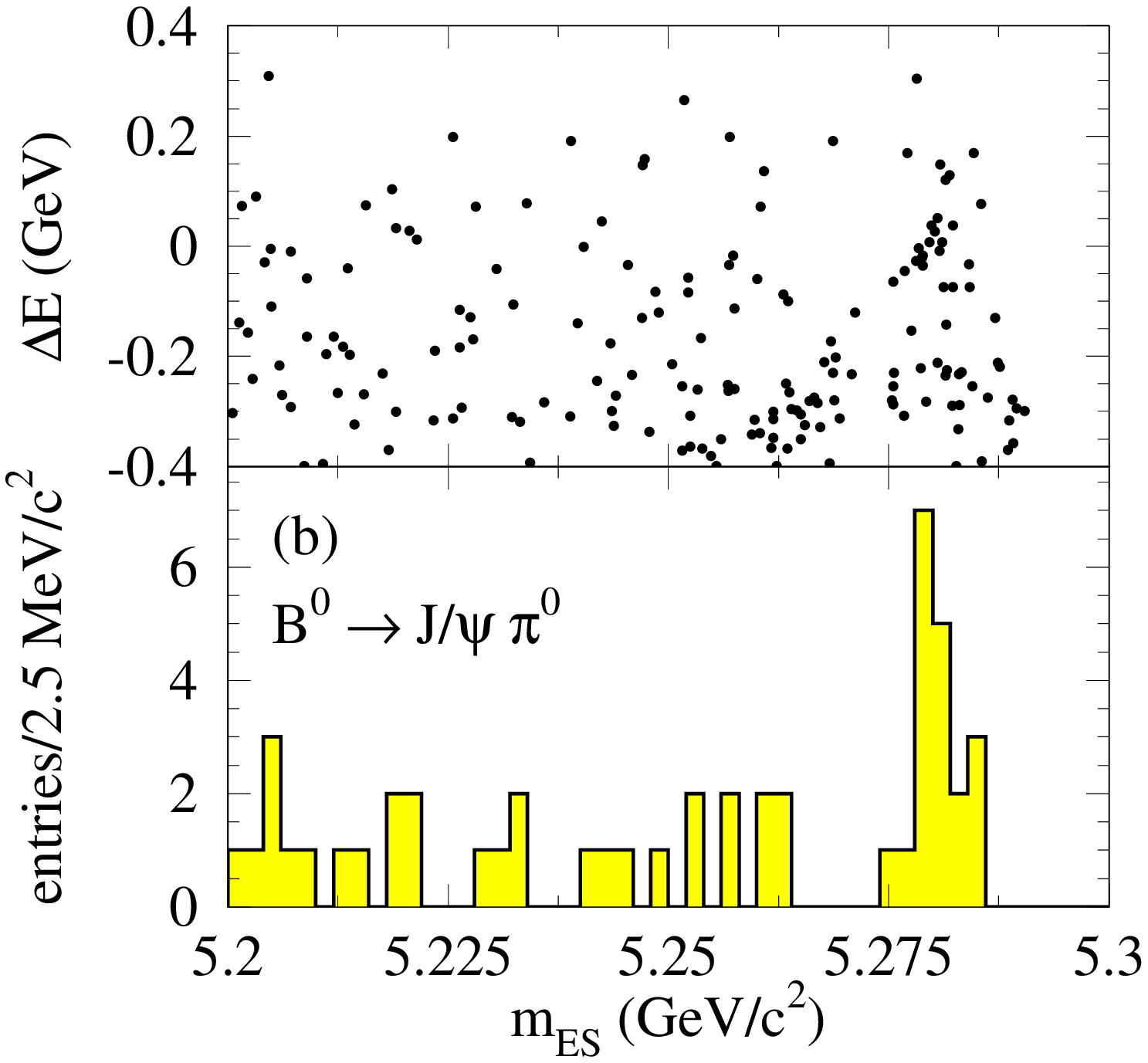,width=0.35\textwidth} &\hspace{-0.25cm} 
       \epsfig{file=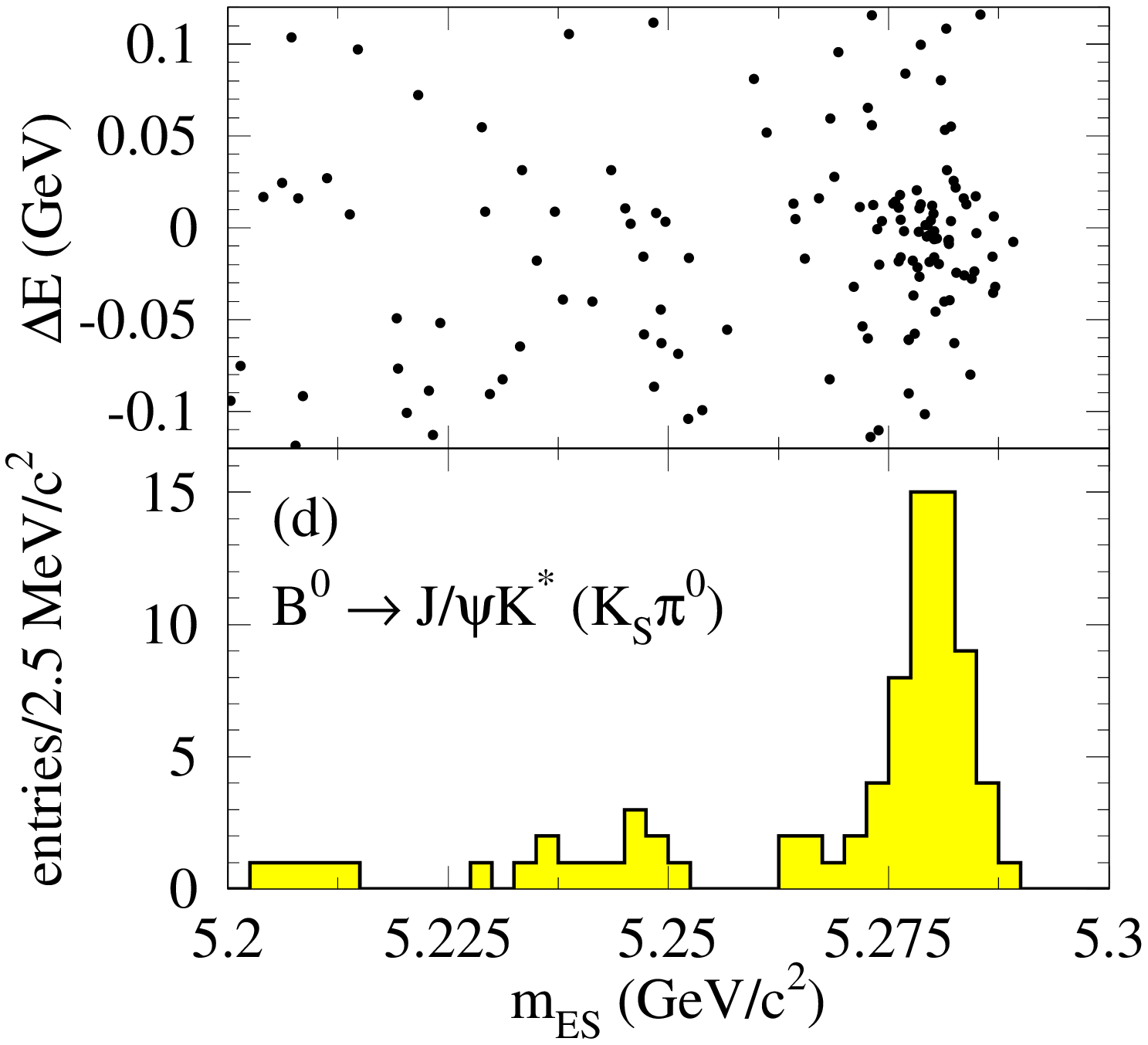,width=0.35\textwidth} &\hspace{-0.25cm} 
	\epsfig{file=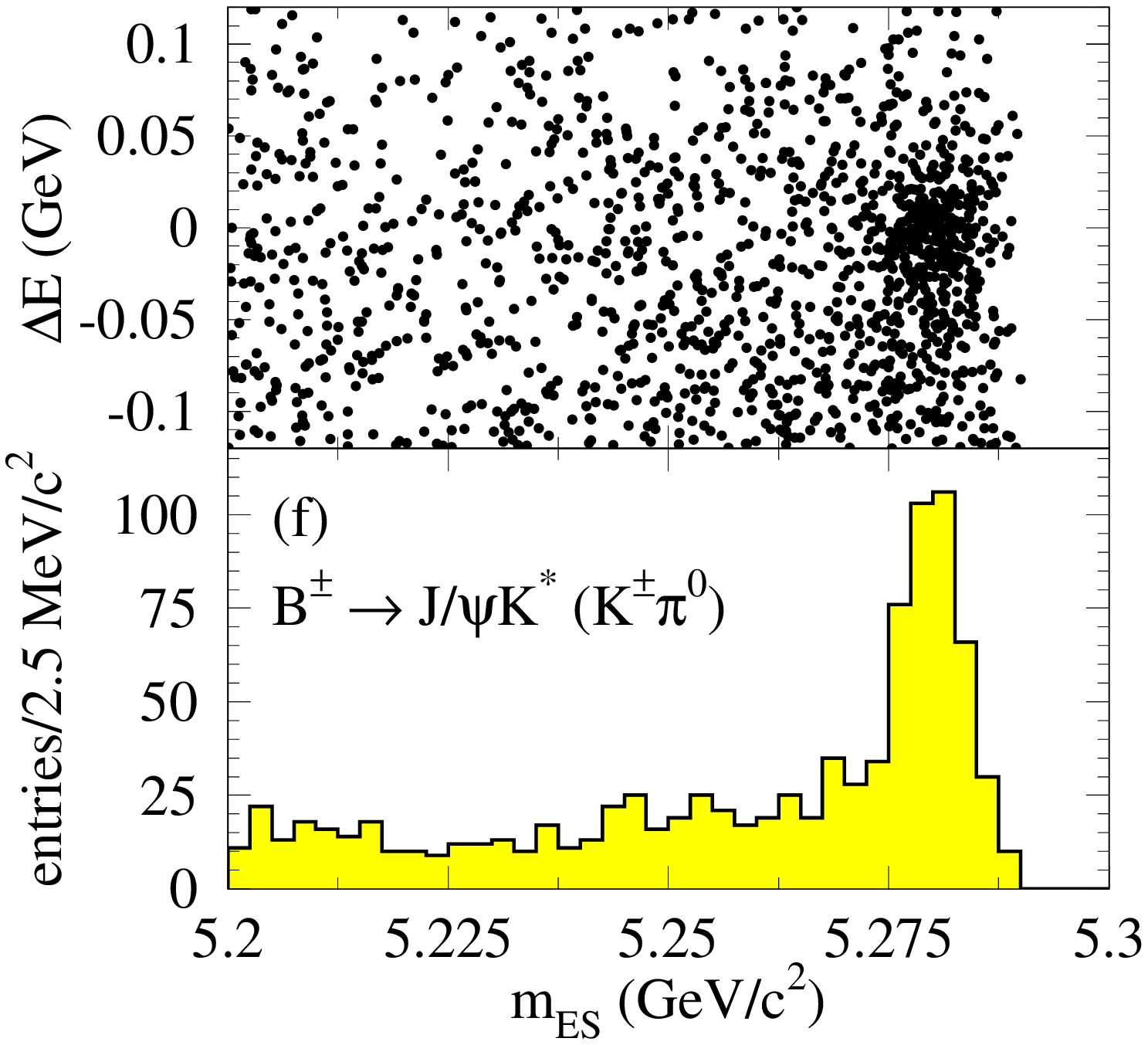,width=0.35\textwidth} 
    \end{tabular}
    \begin{center}
      \caption[fig:OtherJpsi]{
Signal for (a) \bpjpsikp, (b) \bzjpsipiz, (c-d) \bzjpsikstarz, and 
(e-f) \bpjpsikstarp.  The upper plots show the distribution of 
events in the \De-\mes\ plane, and the lower plots show the distribution in
\mes\ of events in the signal region in \De.
\label{fig:OtherJpsi}}
  \end{center}
\end{figure*}

\begin{figure}
\begin{center}
       \epsfig{file=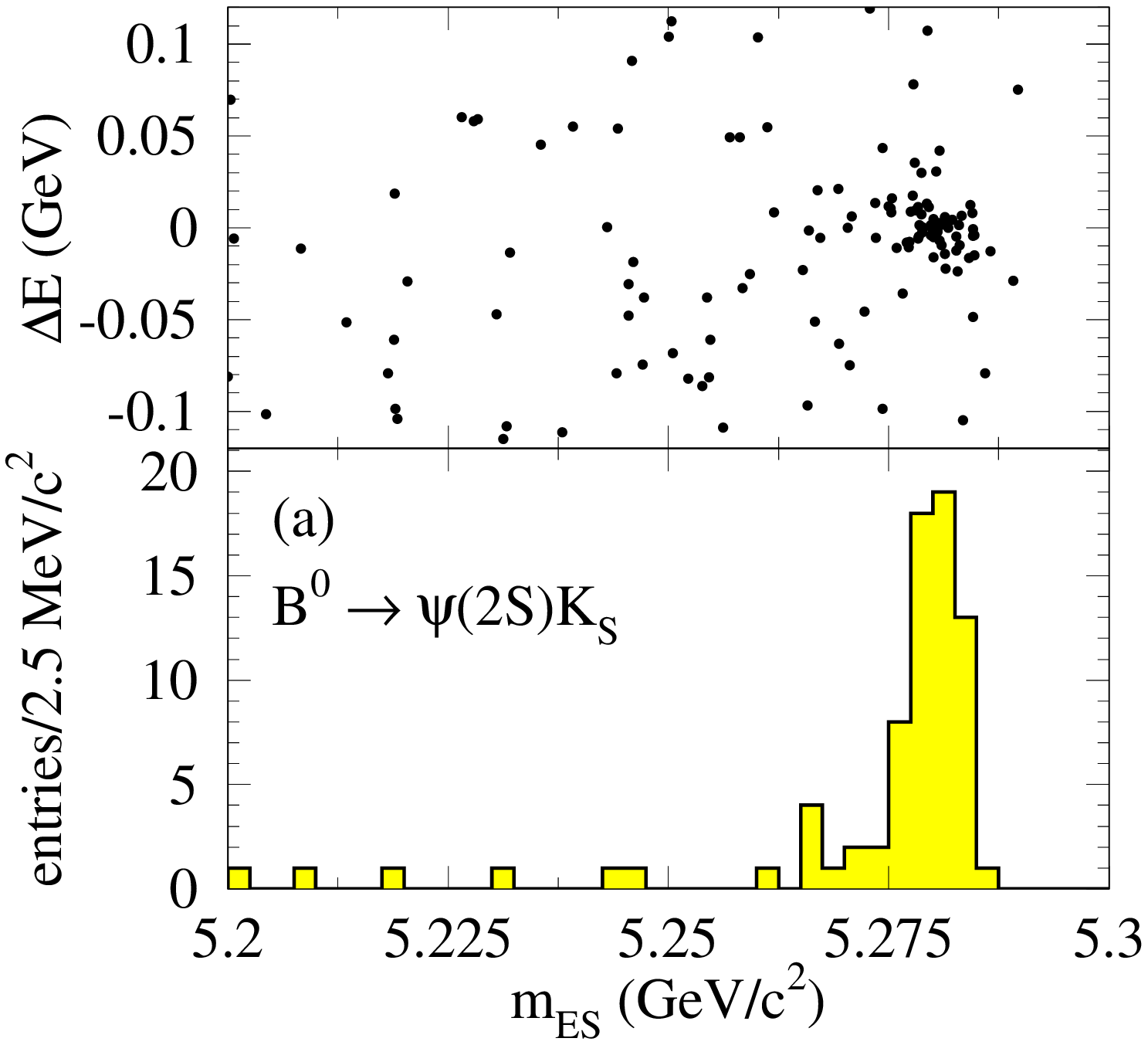,width=0.35\textwidth} \\
       \epsfig{file=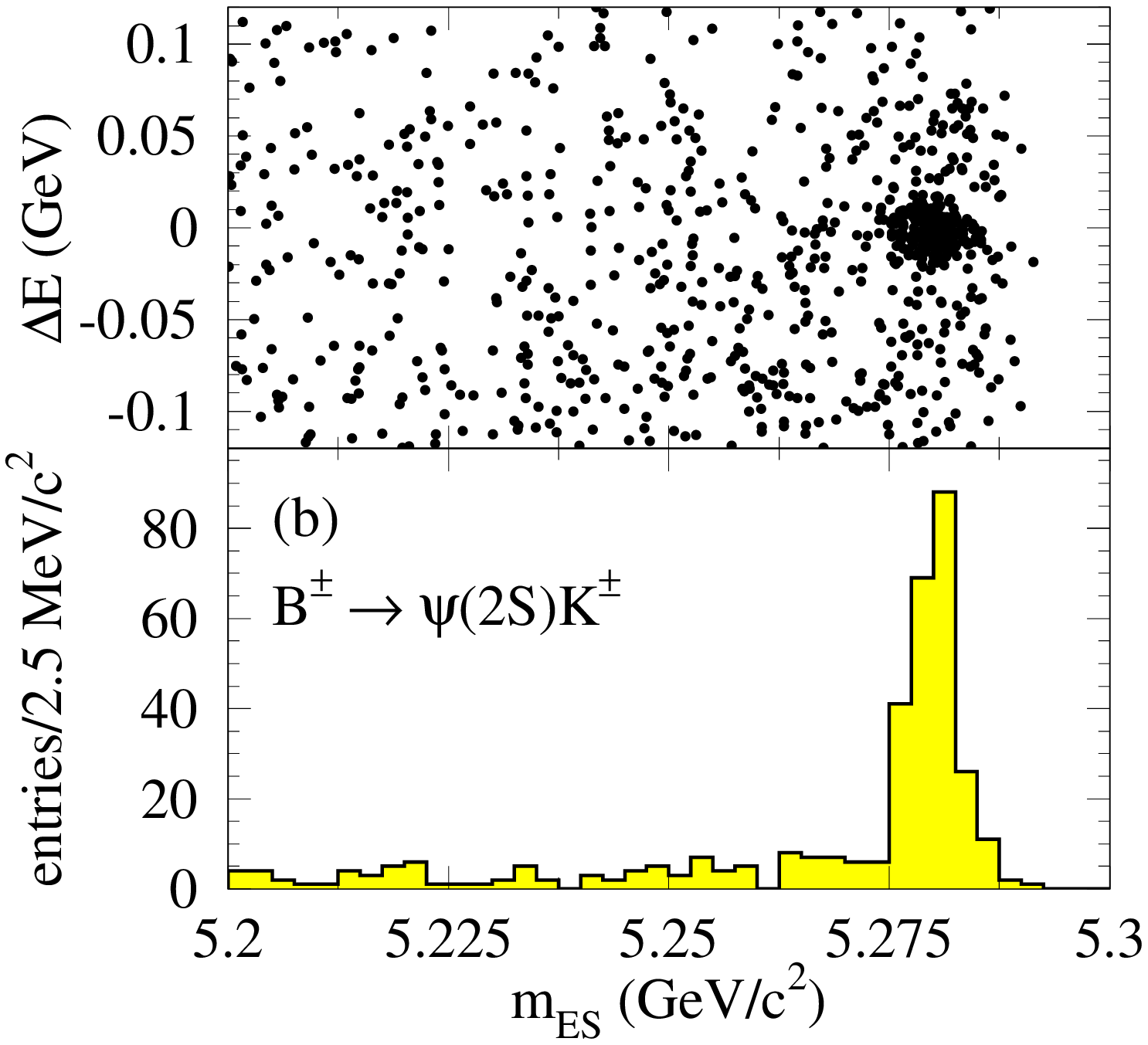,width=0.35\textwidth} 
      \caption[fig:Psi2s]{
Signal for (a) \bzpsitwosks\ and (b) \bppsitwoskp. The upper plots show 
the distribution of 
events in the \De-\mes\ plane, and the lower plots show the distribution in
\mes\ of events in the signal region in \De.
\label{fig:Psi2s}}
  \end{center}    
\end{figure}

\begin{figure}
\begin{center}
       \epsfig{file=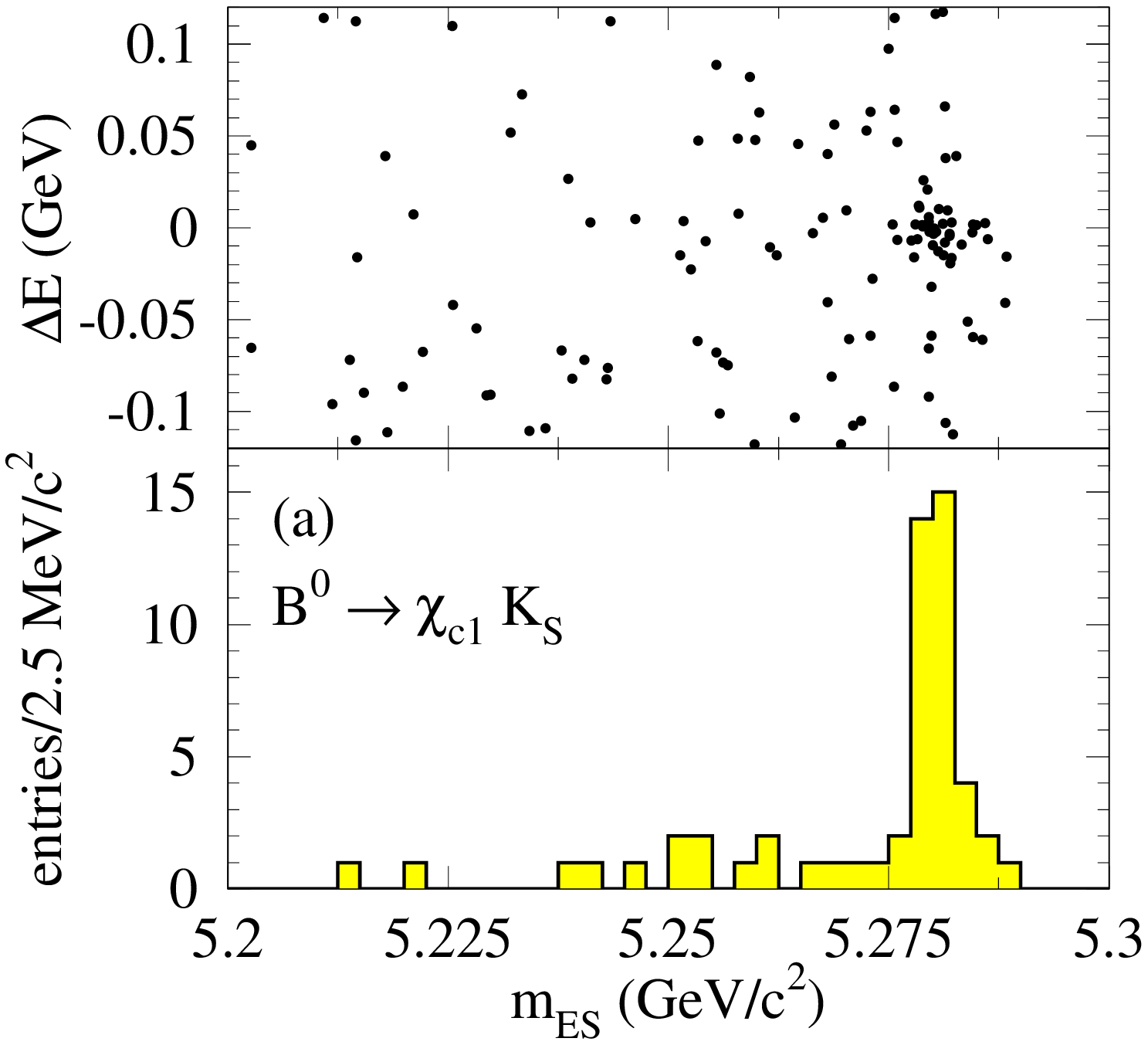,width=0.35\textwidth} \\
       \epsfig{file=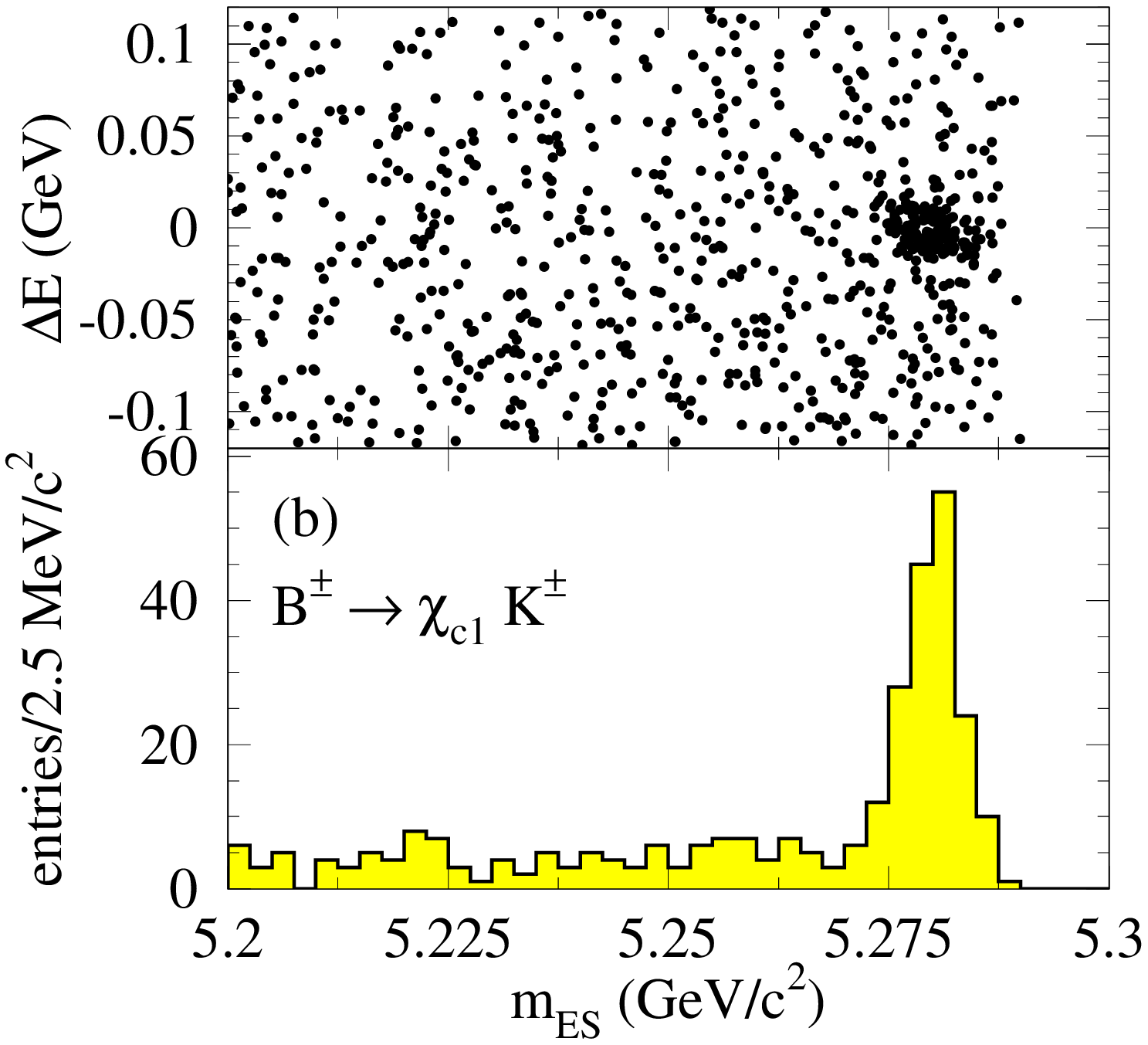,width=0.35\textwidth} \\
       \epsfig{file=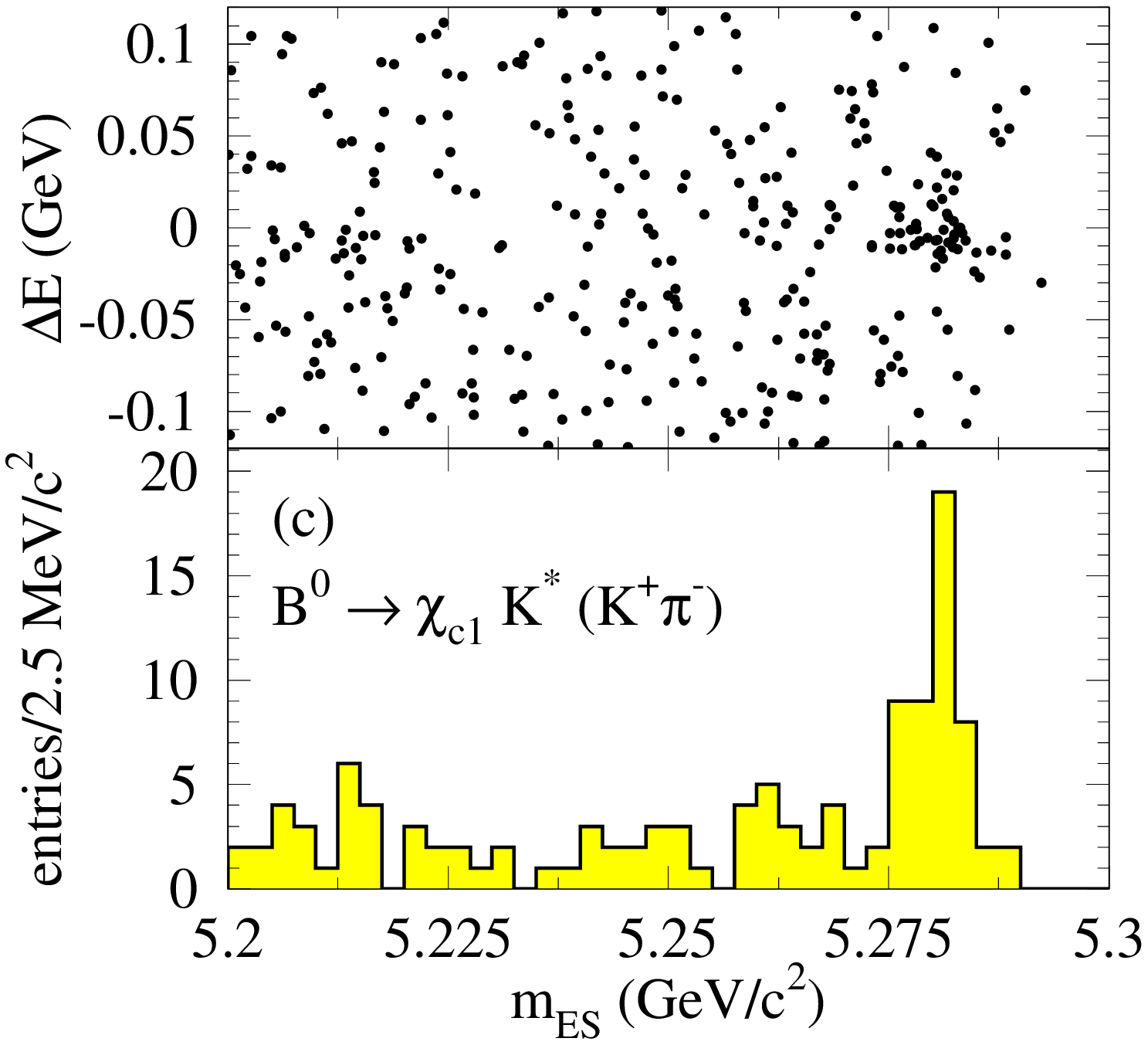,width=0.35\textwidth} 
      \caption[fig:Chic]{
Signal for (a) \bzchiconeks,  (b) \bpchiconekp, and (c) \bzchiconekstarz.
The upper plots show the distribution of 
events in the \De-\mes\ plane, and the lower plots show the distribution in
\mes\ of events in the signal region in \De.
\label{fig:Chic}}
  \end{center}    
\end{figure}

\subsubsection{\bzjpsiks(\pipi)}
All combinations of \jpsi\ and \KS\to\pipi\ candidates are used to 
form $B$ candidates. 
We require the absolute value of $\cos\theta _{\ell}$
to be less than 0.8 and 0.9 for 
\jpsi\to\epem\ and \jpsi\to\mumu\ events respectively.
The distribution of the selected candidates in \De\ and \mes\
 is shown in Fig.~\ref{fig:JpsiK0} (a).

\subsubsection{\bzjpsiks(\ppz)}

All combinations of \jpsi\ and  \KS\to\ppz\ candidates are considered.
For \jpsi\to\epem\ candidates, one track is required to pass the tight or
DCH-only selection, and no particle identification requirement is placed
on the second track.
The mass-constrained \jpsi\ vertex is assumed to be the production point
of the \KS. 
We require that the absolute value 
of  $\cos\theta _{\ell}$ 
 be less 
than 0.7 and 0.8 for \jpsi\to\epem\ and \jpsi\to\mumu\ events respectively. 
The distribution of the selected candidates in \De\ and \mes\ is shown in Fig.~\ref{fig:JpsiK0} (b).
%
%
\subsubsection{ \bpsikl}
Since most of the background in this mode arises from $B$ decays that include
charmonium mesons, we reject events if they contain a candidate for 
\bzjpsiks, \bpjpsikp, \bzjpsikstarz, or \bpjpsikstarp.  The decay modes used
to resconstruct these candidates are the same as those used in the 
branching fraction analysis for each mode, but the selection criteria are
loosened.

Within the remaining events, we select \jpsi\ candidates using a procedure
that differs slightly from the standard
selection.  A vertex constraint is applied, and only candidates for 
which the fit converges are retained.  In addition the momentum of the
\jpsi\ in the center of mass frame is required to
be between 1.4 and 2.0~\gevc, consistent with \bpsikl\ decays.  In the \epem\ mode, one electron candidate is required to
pass the very tight selection and the other the loose selection, and 
the \jpsi\ mass is required to be between 3.00 and 3.13~\gevcc.  For the 
 \mumu\ mode one muon candidate must pass the tight selection and
the other the loose selection, and the \jpsi\ mass is required to be
between 3.06 and 3.13~\gevcc.

We consider all pairs of \KL\ and \jpsi candidates, as
described above, as candidates for $B$ \to $\psi$ \KL\ decays.
We then construct the 
quantity \DeKL\ described previously.

For candidates containing a \KL\ that is identified in the EMC, 
we require that the 
transverse missing momentum be consistent with the 
momentum of the \KL\ candidate calculated from the $B$ mass constraint.  
We compute the missing momentum 
from all tracks and EMC clusters, omitting the \KL\ candidate cluster.  This 
quantity is then projected along the direction of the \KL\ candidate in the 
plane transverse to the beam.  Studies of \bzjpsikl\ events in the 
simulation imply that the event missing momentum should be equal to the 
calculated momentum of the \KL, with a resolution of 0.30 \gevc.  Therefore, we
select 
events where the total missing momentum is not less than 0.65 \gevc\ below the 
calculated \KL\ momentum.  The missing momentum requirement is not applied when
the \KL\ candidate is identified in the IFR, since 
the background is much lower in this sample. 

For all events, we use the angles $\theta _{B}$ and $\theta _{\ell}$ to 
suppress background.    We require that $|\cos\theta_B|$ and
 $|\cos\theta_{\ell}|$ be less 
than 0.9.   To further reduce background,  we also demand that 
$|\cos\theta_B| + |\cos\theta_{\ell}|$  be less than 1.3.

The distribution of the selected candidates in \DeKL\ is shown in
Fig.~\ref{fig:JpsiK0} (c).
%
%
%
%

\subsubsection{\bpjpsikp}  
Every combination of a \jpsi\ candidate and a track is considered.
We require $|\cos\theta _{\ell}|$ to be less than 0.8 and 0.9 for 
\jpsi\to\epem\ and \jpsi\to\mumu\ events respectively.
The distribution of the selected candidates in \De\ and \mes\ is shown in Fig.~\ref{fig:OtherJpsi} (a).
 
%
%
%
\subsubsection{\bzjpsipiz}
For \jpsi\to\mumu\ the standard selection is tightened 
by requiring that
one charged track satisfy the very tight
criteria and the other the loose criteria.
Only  \piz's formed from isolated photon pairs with mass between 
120 and 150 \mevcc\ are considered.

The absolute value of
$\cos\theta_T$
is required to be less than 0.95.  Since background events are 
correlated in $\theta_T$ and $\theta_{\ell}$, we also demand that 
$|\cos\theta_T| + |\cos\theta_{\ell}|$ be less than 1.8.  
The distribution of the selected candidates in \De\ and \mes\ is shown in Fig.~\ref{fig:OtherJpsi} (b).
%
%

\subsubsection{\bzjpsikstarz\ and \bpjpsikstarp}
The \Bz\ 
is reconstructed from pairs of \jpsi\ 
and \Kstarz candidates, while the $B^{+}$ uses \jpsi\ and \Kp\ candidates.
We further require that both
\jpsi\ daughter leptons satisfy either the 
loose muon selection criteria or tight electron selection criteria. 

The distribution of the selected candidates in \De\ and \mes\ are shown 
in Fig.~\ref{fig:OtherJpsi} (c-f). 
%
%
%

\subsubsection{\bzpsitwosks\ and \bppsitwoskp }
Charged $B$ candidates are formed from the combination of a \psitwos\ candidate
with a track, and neutral candidates from the combination
of \psitwos\ and \KS\to\pipi\ candidates.

In the leptonic decay mode of the \psitwos, $|\cos\theta _{\ell}|$
is required to be less than 0.8.  
In the \jpsi\ decay mode of the \psitwos,
$\cos\theta_T$ is required to have an absolute
value of less than 0.9.
The distribution of the selected candidates in \De\ and \mes\ is shown in Fig.~\ref{fig:Psi2s}.
%
%
%
%
\subsubsection{\bzchiconeks\ and \bpchiconekp}
\label{sec:chiks}
\bzchiconeks\ candidates are formed by combining mass-constrained 
\chic{1}\ candidates  
with mass-constrained 
\KS\to\pipi\ candidates. 

\Kp\ candidates are defined  as tracks which lie within the 
angular range $0.35  <  \theta < 2.5$~rad. These are combined with 
mass-constrained  \chic{1} 
candidates to form \bpchiconekp\ candidates. 

The cosine of the $\theta_T$ is required to have 
absolute value less than 0.9. 
The distributions of the selected candidates in \De\ and \mes\ are
shown in Fig.~\ref{fig:Chic} (a-b).

%
%
%

\subsubsection{\bzchiconekstarz} 
$B$ candidates are 
reconstructed by combining mass-constrained \chic{1}\
candidates with \Kstarz\ candidates reconstructed in the $\Kp\pim$ mode. 
We require that the \Kp\ candidate be inconsistent with a pion hypothesis, 
using 
the combined information from \dedx\ measured in the SVT and DCH and 
Cherenkov angle measured in the DIRC. 
We require both tracks from the \jpsi\ to pass either the tight electron 
selection or the loose
muon selection.
\chic{1} candidates are selected if the mass difference between the \chic{1}
 and the 
\jpsi lies between 0.37  and 0.45 \gevcc. 
\Kstarz\ candidates are reconstructed using the standard procedure, 
and are accepted if
the \Kstarz\ mass is within 75 \mevcc\ of the known value~\cite{PDG2000}.

The distribution of the selected candidates in \De\ and \mes\ is shown in Fig.~\ref{fig:Chic} (c).

\section{Background Estimation}

\label{sec:Backgrounds}

\begin{figure}
\begin{center}   
\epsfig{file=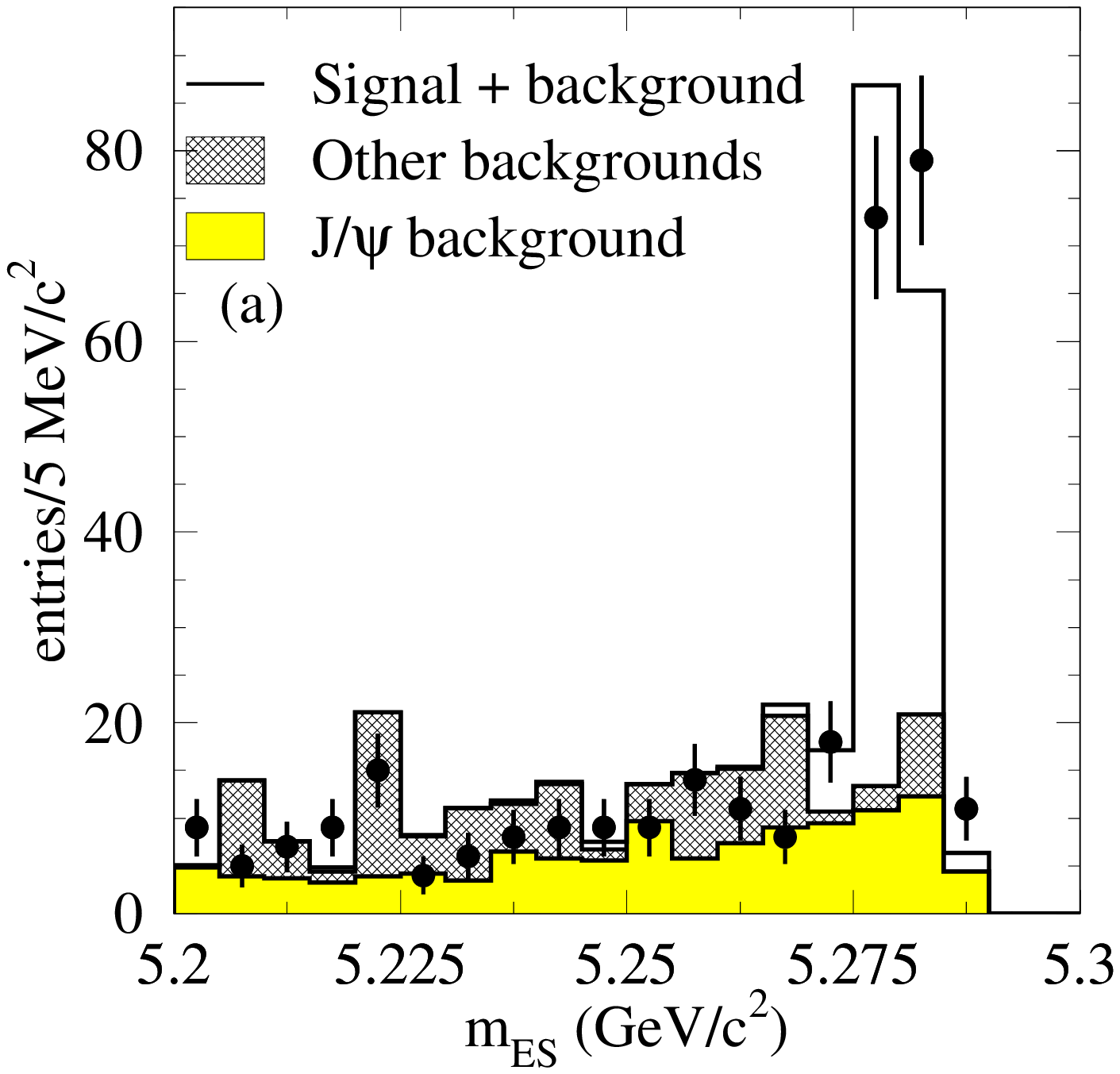,width=0.35\textwidth} 
\epsfig{file=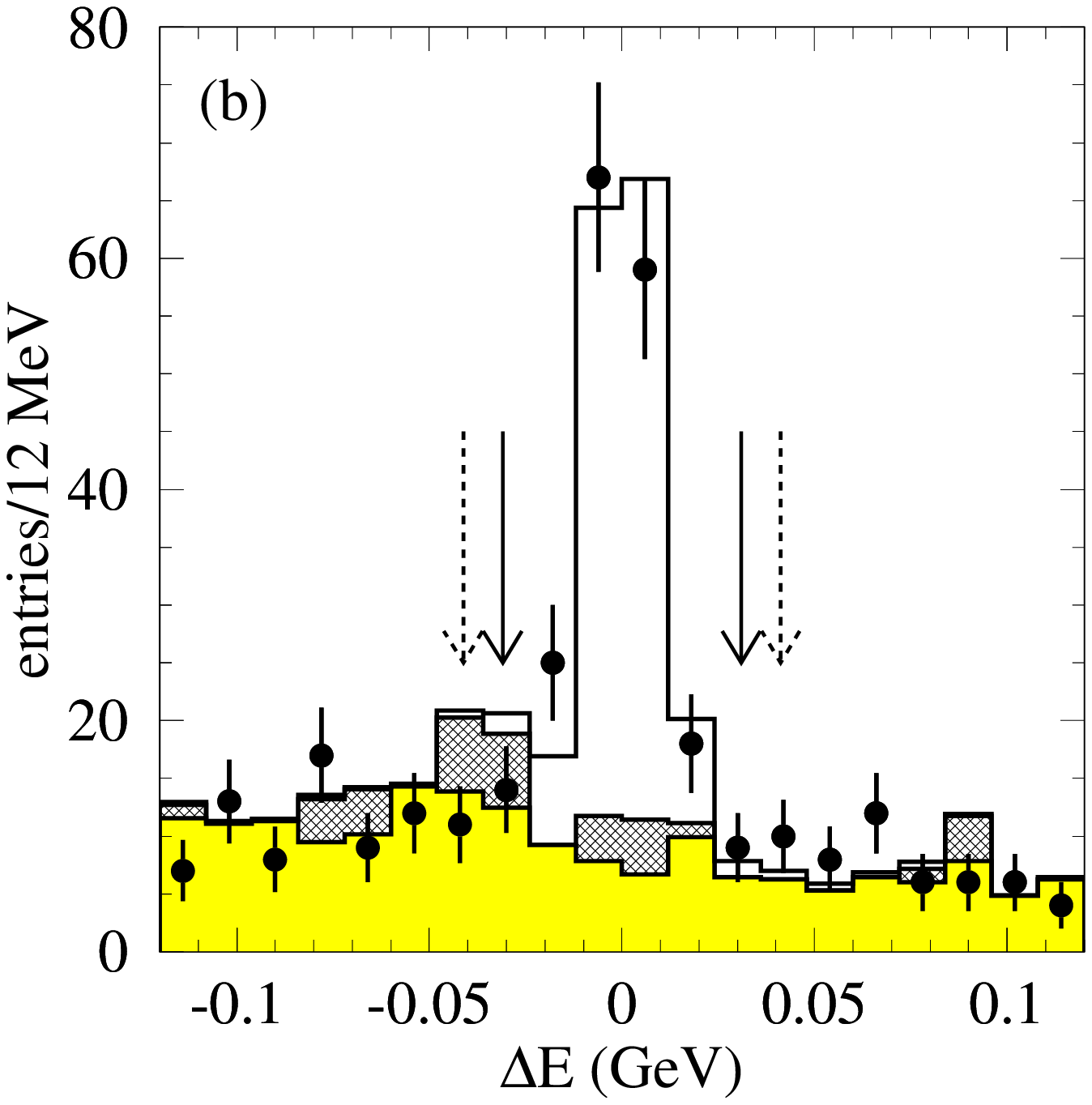,width=0.35\textwidth} 
\end{center}
\caption[fig:incJpsiBkg]{Distribution in (a) \mes\ and (b) \De\ of candidates for
\bpchiconekp.  The points are the data, the shaded histograms are Monte Carlo simulated
background events, broken down into the combinatorial and 
inclusive \jpsi\  contributions, 
and the open histograms are the sum of the Monte Carlo
simulated signal and background distributions.  The Monte Carlo 
distributions are normalized according to the
equivalent luminosity of the samples.  In (b) the \De\ signal region lies 
between the solid arrows, and the sideband region in which we compare the 
observed peaking background to the Monte Carlo prediction lies outside of 
the dashed arrows.  Note that the 
inclusive \jpsi\ background 
peaks in the signal region of \mes, but that neither 
background peaks in the signal region of \De.
\label{fig:incJpsiBkg}}
\end{figure}

\begin{figure}
\begin{center}   
\epsfig{file=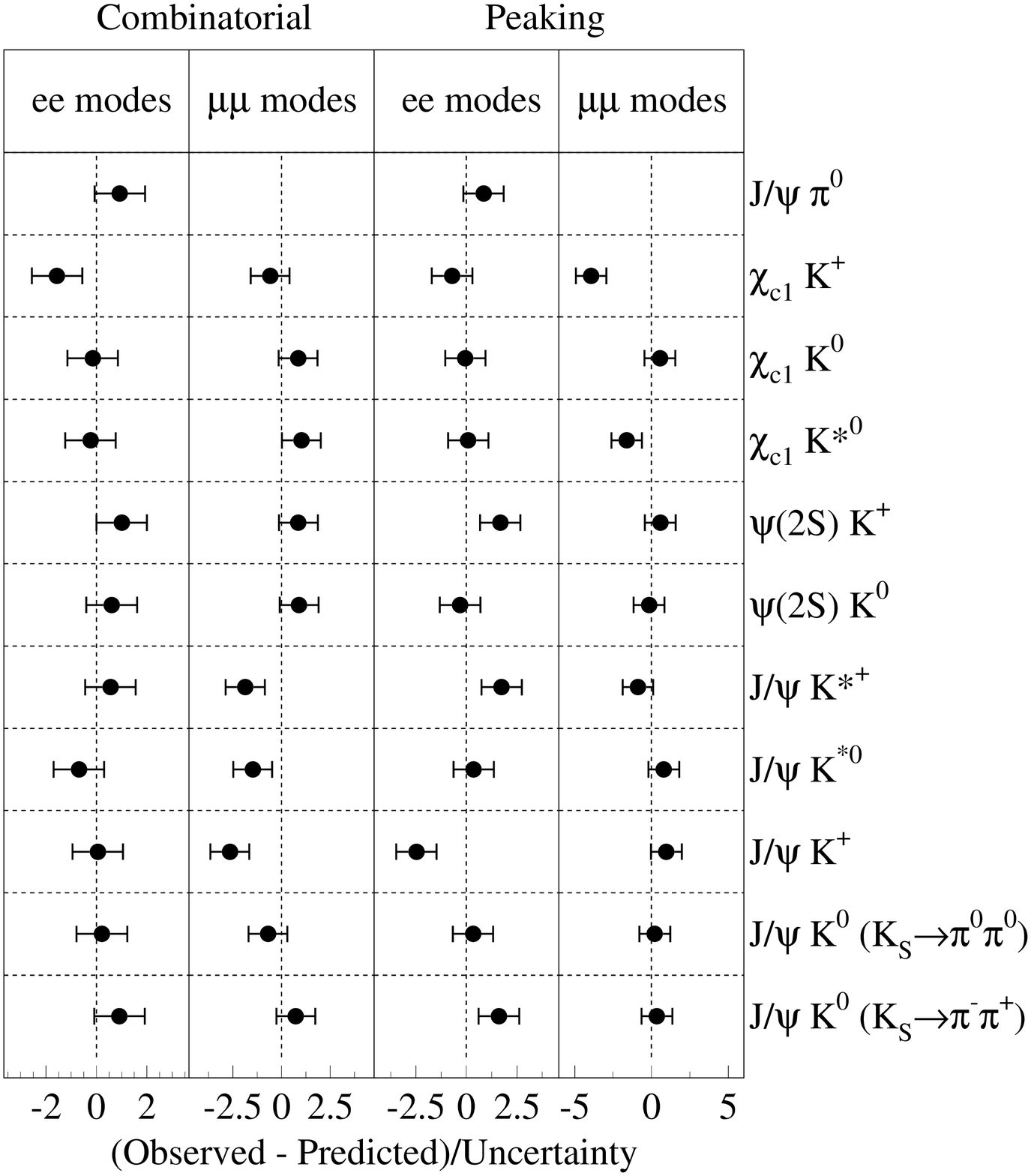,width=0.5\textwidth} 
\end{center}
\caption[debkg]{Difference between the predicted and observed levels of background, divided by the combined statistical error from data and Monte Carlo
simulation.  The comparison of combinatorial 
backgrounds is done in the signal region, while for peaking backgrounds 
the \De\ sideband region is used. For the $\jpsi\piz$ mode the value shown 
is the sum of the
 \epem\ and \mumu\ modes. 
\label{fig:debkg}}
\end{figure}

\begin{table}
\caption{\label{tab:dombkg} Dominant sources of background in the decay modes
we consider, along with the fraction of the total background due to the 
dominant source.  These fractions have substantial uncertainty due to 
the limited statistics of the available Monte Carlo simulation sample.}
\begin{center}
\begin{tabular}{lllc}
\hline \hline
Channel  &   & Dominant   & \% of total \\
         &   & background &             \\ \hline
\bzjpsikz         &  \KS \to \pipi    &  Charmonium & 70 \ \\
                  &  \KS \to \piz\piz &  Continuum \qqbar & 50 \ \\ 
                  &  \KL              &  Charmonium  & 90  \ \\ 
\bpjpsikp         &                   &  Charmonium & 50 \ \\ 
\bzjpsipiz        &                   &  Continuum \qqbar & 55 \ \\ 
\bzjpsikstarz     &                   &  Charmonium & 90   \ \\  
\bpjpsikstarp     &                   &  Charmonium & 85   \ \\  
\bzpsitwoskz      &                   &  Charmonium & 60  \ \\  
\bppsitwoskp      &                   &  Charmonium & 50  \ \\  
\bzchiconekz      &                   &  Charmonium & 95  \ \\  
\bpchiconekp      &                   &  Charmonium & 75  \ \\  
\bzchiconekstarz  &                   &  Charmonium & 90  \ \\  
\hline \hline
\end{tabular}
\end{center}
\end{table}

Backgrounds to the decay modes we measure arise 
predominantly from three sources: 
other $B$ decays that include charmonium mesons in the final state, 
$B$ decays without 
charmonium mesons, and light quark events.
  Monte Carlo simulation studies 
verify that
for $B$ decays without charmonium mesons and for continuum events, 
$B$ candidates
follow the ARGUS distribution in \mes.
On the other hand, the background from inclusive charmonium decays 
includes
modes that are kinematically very similar to the signal modes,
which means that their distribution in \mes\ may have a peak in the signal 
region.    As an example, Fig.~\ref{fig:incJpsiBkg} 
shows the distribution in 
\De\ and \mes\ for signal and background events satisfying the
\bpchiconekp\ selection requirements.  It is critical that 
the so-called ``peaking background'' from other \jpsi\ modes be well 
understood, since it contributes directly as a correction to the fitted 
number of 
signal events in the signal band.

For all modes except \bzjpsikl, we estimate the magnitude of the 
backgrounds by using Monte Carlo simulation, off-resonance data, 
and mass sidebands for \jpsi\ or \psitwos\ candidates in on-resonance data. 
The available Monte Carlo samples are 10 million 
\BB\ decays, the equivalent of 8 \invfb\ of continuum events, and the 
equivalent of several times our data sample
of inclusive $B$ to charmonium decays.

We compare the predicted and
observed levels of background in two regions of the \De-\mes\ plane: the \De\
sideband, defined as that part of the signal neighborhood sufficiently far
from the signal region in \absDe\ that it contains a negligible amount
of signal (typically 4$\sigma$ from zero, though for modes with a \piz\ in the final 
state this is reduced to 3$\sigma$), and the signal 
region.

In each region, we fit a Gaussian and an ARGUS background distribution to the 
observed \mes\ distribution of $B$ candidates in data and Monte Carlo
 samples.  In the \De\ sideband the integral of
 the Gaussian distribution across the \mes\ signal region provides an 
estimate of
 the peaking background.   In the \De\ signal region the integral of the
 ARGUS background function across the \mes\ signal region provides an 
estimate of the combinatorial background. 
A comparison between data and Monte Carlo simulation of the fitted
 results for the combinatorial and peaking background components is 
displayed in Fig.~\ref{fig:debkg}.  In most cases, the
predicted and observed backgrounds are in good agreement, within the 
statistical
errors.  
Discrepancies in the predicted and observed levels of peaking backgrounds in 
the \De\ sideband region are accounted for in our estimation of systematic 
uncertainties.

For the  \bzjpsikl\ sample, 
we estimate the magnitude of the background by 
performing a binned log-likelihood fit to the \DeKL\ distribution in the 
range $-20$ to $80 \mev$.  This fit is described in detail in 
Section~\ref{sec:Physics}.  
The shapes of the signal and inclusive charmonium background components 
are taken from Monte 
Carlo simulation.  The shape of the non-charmonium background component 
is taken from an ARGUS fit
to the \DeKL\ distribution for events in the \jpsi\ mass sideband.  
 To constrain the magnitude of this last component, we first estimate the 
fraction of non-\jpsi\ candidates in the \jpsi\ mass window relative to the 
mass sideband for events with
arbitrary \DeKL. We then scale the number of events with
\DeKL\ between $-20$ and 80 MeV that also have a dilepton invariant mass in
the \jpsi\ sideband region by this fraction to determine the expected 
number of candidates arising from non-charmonium backgrounds.

The dominant source of background for each mode we consider is listed in 
Table~\ref{tab:dombkg}.

\section{Efficiency Calculation}
The selection efficiencies for each mode are 
obtained from detailed Monte Carlo simulations, 
in which the detector response is simulated
using the GEANT3~\cite{GEANT} program. 
In addition, we have used the data where possible to determine the detector 
performance.


We have determined
the efficiency for identifying leptons 
with the sample of inclusively produced \jpsi's in the data.  \jpsi's are
selected by requiring that one track pass the very tight electron or muon selection, with no
lepton identification requirement placed on the other track (the test track). 
The fraction of test
tracks that satisfy a given lepton selection provides a measure of the efficiency for that
selection.


We have determined the 
track finding efficiency from 
multihadron events in the data.
For the standard track selection, the fact that the SVT is an independent
tracking device allows precise determination of the DCH efficiency by
observing the fraction of tracks in the SVT that are also found in the DCH.
For low-momentum pions, such as those produced in the decay 
\psitwos\to\jpsi\pipi, 
\Dstar\ decays are used to
provide information about the efficiency as a function of momentum.  This
measurement takes advantage of the correlation between the pion helicity 
angle in the 
\Dstar\ rest frame and its momentum in the center-of-mass frame.  Since the
helicity angle distribution is known, any deviation between the expected and 
observed distributions can be interpreted as arising from a momentum 
dependence in the track reconstruction efficiency.   
In addition, the efficiency for reconstructing a \KS\to\pipi\ decay
has been determined as a function of the \KS\ flight length
from studies of inclusive \KS\ production in the data.

The efficiency for detecting photon clusters has been determined from the 
data with
a control sample of two-prong \tautau\ events. In the subsample of events tagged by a 
leptonic decay of one of the taus, we compare the number of events with one or two neutral pions, 
and one charged pion, from the second tau decay.
The ratio of these two branching fractions is known to a precision of 1.6\%~\cite{PDG2000}.
By comparing data with simulation, we 
determine a correction factor to be applied to the
photon identification efficiency.  This factor is found to be independent
 of the photon energy.

Both the \jpsi\ mass distribution and \De\ signal distribution in the
\bpjpsikp\ sample have better resolution in 
the simulation than in the data, indicating that the track \pt\ resolution 
in the simulation is overestimated.  To account for this, we
degrade the \pt\ resolution of the simulated tracks by amount chosen
to bring the simulated \jpsi\ mass and \De\ mass distributions into agreement
with those observed in data.


We measure the efficiency of the EMC and the IFR to detect a \KL\
 candidate cluster
 using a control sample of $\epem\to\Phi\gamma$, $\Phi\to\KS\KL$
events.

The efficiencies of the \piz\ veto and missing transverse 
momentum requirements applied for \KL\ reconstruction in the EMC were determined 
using \bpjpsikp\ events.

 The \DeKL\ distribution for simulated  events is
adjusted slightly to account for differences between data and 
Monte Carlo simulation in the beam energy spread
 and \KL\ 
angular resolution.  The correction to the
beam energy spread is derived from a study of \bpjpsikp\ events, and the 
adjustment for the \KL\ angular resolution is determined with the
$\epem\to\Phi\gamma$ control sample. 

The combination of these effects requires a correction
factor 
to be applied to the efficiency determined from the Monte Carlo simulation.
The size of the correction varies among decay modes, and is at most 16\%.

\section{Branching Fraction Determination}
\label{sec:Physics}

\begin{table*}
\caption{\label{tab:syst}
Breakdown of contributions to the systematic errors.
Included are the contributions 
from the secondary branching fractions (${\cal S}$), lepton identification
efficiency (PID), track \pt\ resolution (Trk \pt), track and \KS\to\pipi 
reconstruction efficiency
($\epsilon$(Trk+\KS)), photon identification efficiency ($\epsilon$(\g)), background
determination (BGR), Monte Carlo statistics ($N_{\rm sim}$) and selection 
requirement variation (Sel. var.).
The 1.6\% error from the determination of the number of \BB\ events, which
is common to all modes, is not listed but is included in the total. 
In addition, the statistical uncertainty is shown.  All values are expressed
relative to the measured branching fraction, in percent.
}\medskip
\begin{center}
\begin{tabular}{llcccccccccc}
\hline\hline
Channel & & ${\cal S}$ & PID & Trk \pt & $\epsilon$(Trk+\KS)  & $\epsilon$(\g) & BGR & $N_{\rm sim}$ &Sel. var. & Total & Stat. error\ \\
\hline
\bzjpsikz  & \KS \to \pipi & 1.7  & 1.3 &0.9  & 5.5 & - & 1.1  & 1.3 & 3.5 & 7.3 & 6.4 \\
           & \KS \to \piz\piz & 1.7 & 0.5 & 0.1 &2.4 & 5.0 &2.0 &1.6 & 2.5 & 7.0 & 15.2\\
\bpjpsikp  &  & 1.7 & 1.4 & 1.0 & 3.6 & -  & 1.0 & 0.8 & 2.2  & 5.3 & 3.1\ \\
\bzjpsipiz &  & 1.7 & 2.5 & 0.4 & 2.4 & 2.5 & 1.7 & 1.1 & 10.0 & 11.3 & 32.7\ \\
\bzjpsikstarz &  & 1.7 & 1.3 & 0.8 & 4.7 & 0.2 & 1.4 &0.2 & 4.0 & 6.9 & 4.0\ \\
\bpjpsikstarp &  & 1.7 & 1.3& 1.1 & 4.9 & 1.2 & 2.9 & 0.1 & 5.0 & 8.2 & 6.6 \\
\bzpsitwoskz &  & 9.6 & 1.0 & 1.3 & 7.9 & - & 4.8 & 1.4 & 8.5 & 15.9 & 15.4\ \\
\bppsitwoskp &  & 9.6 & 1.0 & 1.3 & 5.8 & - & 1.3 & 1.6 & 3.7 & 12.1 & 8.1\\
\bzchiconekz &  & 6.2 & 2.4 & 1.2 & 5.6 & 1.3 & 14.5 & 2.2 & 13.2 & 22.0  & 25.1 \\
\bpchiconekp &  & 6.1 & 2.6 & 0.5 & 3.6 & 1.8 & 3.8 & 2.4 & 5.3 & 10.6 & 10.0\\
\bzchiconekstarz &  & 6.2 & 2.4  & 0.8  & 4.8 & 2.7 & 14.3 & 1.8  & 8.1 & 18.7 & 28.8 \\
\hline\hline
\end{tabular}
\end{center}
\end{table*}

\begin{table}
\caption{\label{tab:systkl}
Breakdown of contributions to the systematic error  for the
\bzjpsikl\ analysis.  The statistical error is also shown, with all
values expressed relative to the measured branching fraction, in percent.}\medskip
\begin{center}
\begin{tabular}{lc}
\hline\hline
Source         & Uncertainty  \\
\hline
Tracking efficiency          &  2.4  \\
Lepton identification efficiency            &  1.2  \\
J/$\psi$ mass requirement efficiency &  1.3 \\
$K_L$ efficiency   & 9 \\
$\pi^{0}$ veto efficiency    & 0.7 \\
Missing momentum requirement efficiency   & 0.5 \\
Beam energy  scale (spread)        & 1.0 (3.0) \\
$K_L$ angular resolution     & 4 \\
Branching fractions for B\to\jpsi X  & 3.8 \\
non-$\jpsi$ background shape        & 2 \\
Simulation statistics       & 2.2 \\
Secondary branching fractions  & 1.2 \\
Number of \BB\ events  & 1.6 \\ \hline
Total                  & 12.0 \\ \hline
Statistical error      & 12.0 \\ 
\hline\hline
\end{tabular}
\end{center}
\end{table}

To derive branching fractions 
we have used the secondary branching fractions ${\cal S}$ 
published in Ref.~\cite{PDG2000}.
An exception to this is the branching fraction of \psitwos\to\ellell, 
where we have used our recent measurement of 
$(6.6 \pm 1.1)\times 10^{-3}$~\cite{BabarPub0109}
for the \psitwos\to\mumu mode and the measurement from 
E835~\cite{E835} for the 
\psitwos\to\epem mode.  These measurements are more recent and more accurate 
than those included in Ref. ~\cite{PDG2000}. 
%

We have assumed that \FourS\ decays produce an equal mixture of charged 
and neutral $B$ mesons.
The dependence of our results on this assumption is included 
in Section~\ref{sec:Results}. 

\begin{figure}
    \begin{center}
       \epsfig{file=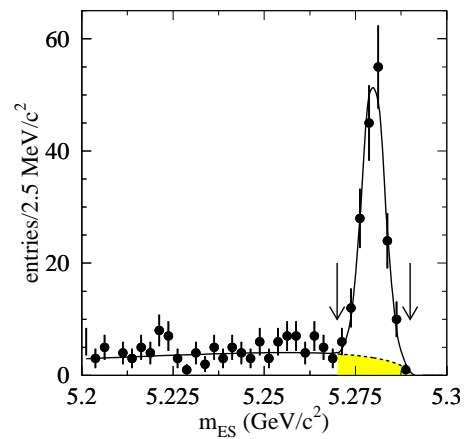,width=0.35\textwidth}   
      \caption[fig:sigCalc]{
Distribution in \mes\ of candidates for \bpchiconekp, with the ARGUS and Gaussian 
fit superimposed.  The number of signal events is calculated by counting the 
events in the 
signal region of \mes (marked by arrows) and subtracting the integral of the fit 
ARGUS function across this region (the shaded portion of the fit) and the peaking
contribution from inclusive \jpsi\ backgrounds, as shown in 
Fig.~\ref{fig:incJpsiBkg}.
\label{fig:sigCalc}}
  \end{center}    
\end{figure}

For all modes except \bzjpsikl, \bzjpsikstarz\ and \bpjpsikstarp,
the number of signal events $N_s$ within the signal region of the 
\De-\mes plane
is determined from the observed number of events after background subtraction.
 The background
has two components, as described in Section~\ref{sec:Backgrounds}:
 a combinatorial component, which is 
obtained by integrating the fitted ARGUS distribution in the signal region, 
and a peaking component that is obtained from inclusive $B\to\jpsi X$ 
simulation after removing the signal channel.  The procedure is illustrated
in Fig.~\ref{fig:sigCalc}.  

We determine the branching
fraction 
by dividing $N_s$ by the 
selection efficiency $\epsilon$, 
${\cal S}$, 
 and the number of \BB\ events in the sample $N_{\BB}$.
Where possible, the branching fraction is determined independently
for the different secondary decay modes, and the results combined
statistically, taking into account correlations in the systematic
errors. For the channels that are statistically
limited, we determine the branching fraction using the combined
sample of candidate $B$ events, irrespective of the secondary decay mode:
\begin{equation}
{\cal B} = {\sum_i N_{s,i} \over N_{\BB} \sum_i \epsilon_i {\cal S}_i}, 
\end{equation} 
where the sum is over all decay modes considered.

The branching fractions for the \bzjpsikstarz\ (${\cal B}^0$) and 
\bpjpsikstarp\  (${\cal B}^+$) modes 
are determined simultaneously from a likelihood fit,  which is
required to account for 
the cross-feed between the \Kstar decay channels.  The cross-feed
is largest for the mode \bpjpsikstarp, where the \Kstarp decays 
to $\Kp\piz$.  In this case, 12\% of the selected candidates arise from
other $B$\to\jpsi\Kstar decays.  The likelihood function includes the
cross-feed contributions as well as all other background sources, 
and has the form:
\begin{equation}
{\cal L}({\cal B}^0,{\cal B}^+) = \prod_{i,j} {\mu_{ij}^{N_{ij}}e^{-\mu_{ij}}
 \over N_{ij}! }
\end{equation}
where $i$ represents a decay mode of the \Kstar\ (to \KS\ or $K^+$), $j$
represents either the  \bzjpsikstarz\ or \bpjpsikstarp\ 
 mode, $N$ is the
observed number of events in the signal region, and $\mu$ is the expected
number of events.  The last is given by:
\begin{equation}
\mu_{ij} = N_{b,ij} + \sum_{i^{\prime}j^{\prime}}{\cal B}^{j^{\prime}}
\epsilon_{iji^{\prime}j^{\prime}}{\cal S}_{i^{\prime}j^{\prime}}N_{\BB}
\end{equation}
where $N_b$ is the number of background events estimated in the same manner as
for the other channels. The four indices attached to the selection
efficiencies denote the fraction of events in the $i^{\prime}j^{\prime}$
 mode that
pass the $ij$ selection requirements, as determined with the Monte Carlo
simulation. 

We determine the number of signal  and background events for the 
\bzjpsikl\ decay mode by performing a binned 
likelihood fit to the \DeKL\ distribution.  The fit takes as input
$a_i$, the fraction of simulated \bzjpsikl\ events in the $i^{\rm th}$ bin, 
$b_i$, the fraction
of simulated inclusive charmonium background events in the $i^{\rm th}$ bin, 
$c_i$, 
the fraction of non-charmonium background events from the 
mass sidebands of the \jpsi\ in the $i^{\rm th}$ bin, and $d_i$,
the number of data events in the $i^{\rm th}$ bin.  The likelihood function has the 
form:
\begin{eqnarray}
{\cal L}(N_s, N_{\psi X}, N_{{\rm non-}\psi}) &=& \prod_{i=1}^{N_{\rm bin}} { \mu_i^{d_i} e^{-\mu_i} \over d_i !} \\
\nonumber        &  & \times {e^{-(N_{{\rm non-}\psi} -M)^2 \over 2 \left(\sigma^2 + N_{{\rm non-}\psi} \right)} \over 
\sqrt{2\pi \left(\sigma^2 +N_{{\rm non-}\psi} \right) }} 
\end{eqnarray}
where $N_{\psi X}$ is the number of inclusive charmonium background events, 
$N_{{\rm non-}\psi}$ is the number of non-charmonium background events, $M$ 
is the expected number of non-charmonium background events determined from the 
mass sidebands of the \jpsi, $\sigma$ is the uncertainty on $M$, and $\mu_i$
 is the
expected number of events in the $i^{\rm th}$ bin, defined as:
\begin{equation}
\mu_i \equiv N_s a_i + N_{\psi X} b_i + N_{{\rm non-}\psi} c_i
\end{equation}

\section{Systematic Errors}

Systematic errors on the results arise from the uncertainty on the number of
 \BB\ events, the secondary branching
fractions of the modes considered, the estimate of the selection
efficiency and the knowledge of the background level. 
The size of the various contributions to the systematic error,
expressed as a fraction of the branching fraction value, is listed for all modes
except \bzjpsikl\ in Table~\ref{tab:syst} and for the \bzjpsikl\ mode in 
Table~\ref{tab:systkl}.

The uncertainty on the number of \BB\ events introduces a systematic
error of 1.6\% in common for all modes. 
The uncertainties in the branching fractions of the secondary decay modes
lead to a systematic error of between
1.7\% and 9.8\%, depending on the mode considered.

The systematic error due to the finite size of the 
available Monte Carlo sample
is between 0.1\% and 2.4\% for the different modes.

We have determined the efficiency for a charged particle to be reconstructed as a track
that passes the standard track selection
to a precision of 1.2\% per track. 
The uncertainty in the reconstruction efficiency for the 
low-momentum pions from the $\psitwos\to\jpsi\pipi$ decay is determined
to be 2\% per track.  The systematic error associated with
reconstructing a \KS\to\pipi decay has two sources: 
knowledge of the reconstruction
efficiency
for the two $\pi$ tracks, and differences in the selection criteria 
efficiencies observed between the inclusive \KS\ data and the Monte Carlo
simulation. 
The observed discrepancies and their statistical uncertainties are
summed in quadrature to yield a systematic error of approximately 5\%.

The systematic error on lepton identification efficiencies arise from 
the statistics of the inclusive \jpsi\ sample, and 
from comparing the efficiencies in different low-multiplicity control 
samples. It
varies from 0.5\% to 2.8\% per \jpsi\ or \psitwos\ depending 
on the criteria used to select the leptons.

The quality of the simulation of photon detection and energy measurement in 
the EMC has been validated by a detailed comparison between real and simulated 
data. In particular, the position and resolution of the \piz\ and $\eta$ 
mass peaks 
in the photon pair mass spectrum has been compared as function of photon 
energy, calorimeter occupancy and time of data collection. The agreement in 
terms of energy scale is found to be better than 0.75\% in all cases; energy 
resolution is also well described at the level of 1.5\%. The absolute photon 
detection efficiency is known to 1.25\%. The resulting 
systematic errors on the branching fractions are in the range of 
1.3\% to 5\% depending on the decay mode.

We account for the uncertainty in the \pt\ resolution by varying the
amount by which the Monte Carlo simulated momentum resolution is 
degraded within the
range in which the data and Monte Carlo \jpsi mass and \De\ widths are
compatible.  The observed variation in selection efficiency is between 
0.1\% and 1.3\%. To account for the possibility that other variables used
in selecting candidates may not be perfectly modelled in the simulation, 
we vary the
selection requirements 
and repeat the branching fraction measurement.  In most cases the range
of variation is $\pm 1 \sigma$, where $\sigma$ is the width observed in data
for the variable under consideration, while for helicity angles a variation
of $\pm 0.05$ in their cosine is used.  
The observed 
variations in the results are between 2.5\% and 14.1\%. 
Modes 
with a \Kstar\ in the final state merit special mention, since there can 
be some variation of
selection efficiency with the polarization of the vector meson, and 
the polarization amplitudes are subject to 
experimental uncertainty. 
The Monte Carlo
simulation from which we derive our efficiency assumes the
polarization amplitudes measured by CLEO~\cite{CLEOamp}.  We have studied the
changes in efficiency that occur when
the amplitudes are varied by twice the difference between the values
measured at CLEO and \babar~\cite{AngAnalysis}.  We find that these
changes are consistent with those observed when the selection
requirement on $\theta_K$ is varied.  

For the \bzjpsikl\ analysis, we include additional systematic errors 
associated with the selection efficiency. These originate from the uncertainty 
in the \KL\
reconstruction efficiency 
and angular resolution determined from data, the knowledge 
of the absolute scale and spread of the beam energy, 
and from the various selection requirements used to isolate the signal.


Another systematic error 
arises from our knowledge of the backgrounds.
For all modes except \bzjpsikl, we use data in the \De\ sideband to estimate this
uncertainty.  
We determine the uncertainty in the size of  
the combinatorial background by 
repeating the fit to the data with the shape of the
ARGUS function (the parameter $c$ in Equation~\ref{eq:ARGUS}) fixed to the value 
obtained from fitting the \De\ sidebands, 
allowing only the normalization to vary.  This accounts for any correlation
between the ARGUS and Gaussian fits in the \De\ signal region.
We estimate the uncertainty in the predicted size of the peaking background 
by comparing the observed Gaussian component in the 
\De\ sideband to that estimated from the inclusive 
$B\to\jpsi X$ simulation.  This procedure takes advantage of the fact 
that the distribution of candidates from this background in
\De\ depends primarily on kinematics rather than the poorly-known composition
of the background.  In particular, the background does not peak in the
signal region of \De\ (see Fig.~\ref{fig:incJpsiBkg}), which implies that the
relative normalization observed in the \De\ sideband can also be expected to
hold in the signal region. 
The systematic error attributed to the knowledge of the backgrounds
varies from
1.0\% to 14.5\% for the various modes. 
In addition, for the \bzjpsikstarz,
\bpjpsikstarp\ and \bzchiconekstarz\ modes,
a systematic error is included to account for
the uncertainty in the non-resonant $B\to\jpsi K\pi$ branching fractions, and
the contribution of feed-down from higher \Kstar\ resonances.
This ranges from 1.4\% to
3.7\% depending on the mode.

For the \bzjpsikl\ decay mode, we determine
the uncertainty arising from knowledge of the  shape of 
the non-\jpsi\ background both by changing the 
fitted parameters of the ARGUS function for this background component
by one standard deviation and 
also
directly in the fit by using the \DeKL\ distribution from the non-\jpsi\ events in the
data. 
The analysis is also repeated after varying the values of the 
 branching fractions for the component modes
in the simulation of $B\to\jpsi X$ decays by
the uncertainty quoted in Ref.~\cite{PDG2000}. This is done separately
for the main background modes and then for all the remaining modes
together. 
Since the non-resonant $B\to\jpsi K \pi$ component is poorly measured, we
vary it in the range from -50\% to +400\%. 

\section{Results}
\label{sec:Results}

\begin{table}
\caption{\label{tab:results}Measured branching fractions
for exclusive decays of $B$ mesons involving charmonium.  
The first error is statistical and the second systematic.}
\begin{center}
\begin{tabular}{llccccl}
\hline \hline
Channel &   & \multicolumn{5}{c}{Branching fraction/$10^{-4}$}\\
\hline
\bzjpsikz         &  \KS \to \pipi    & 8.5&$\pm$&0.5&$\pm$&0.6 \ \\
                  &  \KS \to \piz\piz & 9.6&$\pm$&1.5&$\pm$&0.7  \ \\ 
                  &  \KL              & 6.8&$\pm$&0.8&$\pm$&0.8   \ \\ 
                  &  All              & 8.3&$\pm$&0.4&$\pm$&0.5  \ \\ 
\bpjpsikp         &                   & 10.1&$\pm$&0.3&$\pm$&0.5 \ \\ 
\bzjpsipiz        &                   & 0.20&$\pm$&0.06&$\pm$&0.02 \ \\ 
\bzjpsikstarz     &                   & 12.4&$\pm$&0.5&$\pm$&0.9   \ \\  
\bpjpsikstarp     &                   & 13.7&$\pm$&0.9&$\pm$&1.1   \ \\  
\bzpsitwoskz      &                   & 6.9&$\pm$&1.1&$\pm$&1.1  \ \\  
\bppsitwoskp      &                   & 6.4&$\pm$&0.5&$\pm$&0.8  \ \\  
\bzchiconekz      &                   & 5.4&$\pm$&1.4&$\pm$&1.1  \ \\  
\bpchiconekp      &                   & 7.5&$\pm$&0.8&$\pm$&0.8  \ \\  
\bzchiconekstarz  &                   & 4.8&$\pm$&1.4&$\pm$&0.9   \ \\  
\hline \hline
\end{tabular}
\end{center}
\end{table}

In Table~\ref{tab:results} we summarize our branching fraction measurements.
The observed number of events in the signal region, 
the predicted background, and the selection efficiency are given in
Table~\ref{tab:yield}. 

\begin{table*}
\caption{\label{tab:yield}The observed number of events
in the signal region, estimated background, 
efficiency, efficiency times secondary branching fractions
and measured branching fraction
for exclusive decays of $B$ mesons involving charmonium.  The combinatorial
background is estimated from a fit to the signal plus sideband region
in \mes, while the peaking background is estimated with
Monte Carlo.  For the \bzjpsikl\ mode the inclusive charmonium background
is listed in the ``Peaking'' column and the other backgrounds in the
``Combinatorial'' column.
For the branching fractions, 
the first error is statistical and the second systematic.}\medskip
\begin{center}
\begin{tabular}{llrrclrclccccccl}
\hline\hline
Channel &   & $N_{\rm obs}$  & \multicolumn{3}{c}{Combinatorial Bkgr} & \multicolumn{3}{c}{Peaking Bkgr} & Efficiency (\%) & Eff $\times {\cal S}$(\%) & \multicolumn{5}{c}{Branching fraction/$10^{-4}$}\\
\hline
\bzjpsikz         &  \KS \to \pipi    & 275  & 6.1 &$\pm$& 2.7  & 3.4&$\pm$&1.1 & 33.8 & 1.37 & 8.5&$\pm$&0.5&$\pm$&0.6 \ \\
                  &  \KS \to \piz\piz & 77  & 12.2&$\pm$&3.7  & 2.3&$\pm$&0.9 &15.5 & 0.29 & 9.6&$\pm$&1.5&$\pm$&0.7  \ \\ 
                  &  \KL              & 408  & 25&$\pm$&3  & 200&$\pm$&14 & 22.3  & 1.46& 6.8&$\pm$&0.8&$\pm$&0.8   \ \\ 
                  &  All              &   & & &  & & & &  &  & 8.3&$\pm$&0.4&$\pm$&0.5  \ \\ 
\bpjpsikp         &                   & 1135  & 8.9&$\pm$&2.6  & 17.1&$\pm$&2.6 & 41.2 & 4.86 & 10.1&$\pm$&0.3&$\pm$&0.5 \ \\ 
\bzjpsipiz        &                   & 19  & 4.7&$\pm$&0.9  &0.7&$\pm$&0.1  & 25.8 &  3.01 & 0.20&$\pm$&0.06&$\pm$&0.02 \ \\ 
\bzjpsikstarz     &                   &  695 & 50.2&$\pm$&7.8  & 50.0&$\pm$&3.3 & 22.6 & 1.10 &12.4&$\pm$&0.5&$\pm$&0.9   \ \\  
\bpjpsikstarp     &                   &  625 & 160.6&$\pm$&15.9 & 87.0&$\pm$&5.8 & 17.9 & 1.09 & 13.7&$\pm$&0.9&$\pm$&1.1   \ \\  
\bzpsitwoskz      &                   & 63  &  6.0&$\pm$&3.3 & 1.0&$\pm$& 0.8 & 22.0 & 0.37& 6.9&$\pm$&1.1&$\pm$&1.1  \ \\  
\bppsitwoskp      &                   & 247 & 27.2&$\pm$&5.5  & 12.5&$\pm$&2.8 & 29.6 & 1.46 & 6.4&$\pm$&0.5&$\pm$&0.8  \ \\  
\bzchiconekz      &                   & 37 & 7.2&$\pm$&2.1  & 3.7&$\pm$&1.3 & 19.1 & 0.21& 5.4&$\pm$&1.4&$\pm$&1.1  \ \\  
\bpchiconekp      &                   & 179  & 24.2&$\pm$&4.7  &9.7&$\pm$&2.7 & 26.3  & 0.85 & 7.5&$\pm$&0.8&$\pm$&0.8  \ \\  
\bzchiconekstarz  &                   & 52  & 13.0&$\pm$&1.6  & 6.4&$\pm$&5.8 & 13.9 & 0.30& 4.8&$\pm$&1.4&$\pm$&0.9   \ \\  
\hline\hline
\end{tabular}
\end{center}
\end{table*}


From these results, we have determined the following ratios of charged to
neutral branching fractions, where the first error is statistical and the
second systematic:

\begin{eqnarray}
\frac{{\cal B}(\bpjpsikp)}{{\cal B}(\bzjpsikz)} & = & 1.20 \pm 0.07 \pm 0.04\\
\frac{{\cal B}(\bpjpsikstarp)}{{\cal B}(\bzjpsikstarz)} & = & 1.10 \pm 0.09 \pm 0.08\\
\frac{{\cal B}(\bppsitwoskp)}{{\cal B}(\bzpsitwoskz)} & = & 0.94 \pm 0.16 \pm 0.10\\
\frac{{\cal B}(\bpchiconekp)}{{\cal B}(\bzchiconekz)} & = & 1.39 \pm 0.37 \pm 0.22
 \end{eqnarray}
Combining all of these measurements yields:
\begin{eqnarray}
\frac{{\cal B}(B^+\to {\rm Charmonium})}{{\cal B}(\Bz\to {\rm Charmonium})} &=& 1.17 \pm 0.07 \pm 0.04
 \end{eqnarray}

Assuming equal partial widths for $\Bz\to\jpsi h^0$ and $B^+\to\jpsi h^+$ for 
any meson $h$ and 
using the known ratio of the
charged to neutral $B$ meson
lifetimes $\tau_{B^+}/\tau_{B^0} = 1.062 \pm 0.029$~\cite{PDG2000},
we find:
\begin{equation}
R^{+/0} \equiv \frac{{\cal B}(\FourS\to\Bu\Bub)}{{\cal B}(\FourS\to\Bz\Bzb)} 
= 1.10 \pm 0.06 \pm 0.05
\end{equation}

We provide the formulae for recomputing our results for an arbitrary value of 
$R^{+/0}$, rather than the value of unity we have assumed:
\begin{eqnarray}
{\cal B}(\Bu\to X,R^{+/0}) &=& \frac{(1+R^{+/0})}{2R^{+/0}} {\cal B}(\Bu\to X, 1)  \\
{\cal B}(\Bz\to X,R^{+/0}) &=& \frac{(1+R^{+/0})}{2} {\cal B}(\Bz\to X, 1)  
\end{eqnarray}

We also determine the ratio of branching fractions for a vector versus scalar
light meson accompanying the charmonium meson:

\begin{eqnarray}
\frac{{\cal B}(\bzjpsikstarz)}{{\cal B}(\bzjpsikz)} & = & 1.49 \pm 0.10 \pm 0.08 \\
\frac{{\cal B}(\bpjpsikstarp)}{{\cal B}(\bpjpsikp)} & = & 1.37 \pm 0.10 \pm 0.08 \\
\frac{{\cal B}(\bzchiconekstarz)}{{\cal B}(\bzchiconekz)} & = & 0.89 \pm 0.34 \pm 0.17 
\end{eqnarray}
These three ratios are consistent and yield an average value:
\begin{equation}
\frac{{\cal B}(B\to {\rm charmonium+vector})}{{\cal B}(B\to {\rm charmonium+scalar})} =
1.40 \pm 0.07 \pm 0.06
\end{equation}

Finally, the following ratios between the production rates for different charmonium states 
have been determined:
\begin{eqnarray}
\frac{{\cal B}(\bzpsitwoskz)}{{\cal B}(\bzjpsikz)}&=&0.82 \pm 0.13 \pm 0.12 \\
\frac{{\cal B}(\bzchiconekz)}{{\cal B}(\bzjpsikz)}&=&0.66 \pm 0.11 \pm 0.17 \\
\frac{{\cal B}(\bppsitwoskp)}{{\cal B}(\bpjpsikp)}&=&0.64 \pm 0.06 \pm 0.07 \\
\frac{{\cal B}(\bpchiconekp)}{{\cal B}(\bpjpsikp)}&=&0.75 \pm 0.08 \pm 0.05 
\end{eqnarray}



\section{Summary}
\label{sec:Summary}

We have presented
measurements of branching fractions of $B$ mesons to several
two-body final states that include a \jpsi,\psitwos\ or \chicone\ meson and 
a \Kz, \Kp, \Kstar or \piz. 
Our results are in good agreement with previous measurements
\cite{PDG2000} and have superior 
precision, both in terms of individual branching fractions and their ratios. 
In addition, based on isospin invariance, we find the ratio 
of charged to neutral $B$ meson production on the \FourS\ resonance to be 
compatible 
with unity within two standard deviations, and also compatible with
the measurement reported by CLEO~\cite{ref:cleoR}.  Our central
value and CLEO's are both higher than one, with the difference in our case 
larger than one standard deviation.

\section{Acknowledgments}
\label{sec:Acknowledgments}

We are grateful for the 
extraordinary contributions of our \pep2\ colleagues in
achieving the excellent luminosity and machine conditions
that have made this work possible.
The collaborating institutions wish to thank 
SLAC for its support and the kind hospitality extended to them. 
This work is supported by the
US Department of Energy
and National Science Foundation, the
Natural Sciences and Engineering Research Council (Canada),
Institute of High Energy Physics (China), the
Commissariat \`a l'Energie Atomique and
Institut National de Physique Nucl\'eaire et de Physique des Particules
(France), the
Bundesministerium f\"ur Bildung und Forschung
(Germany), the
Istituto Nazionale di Fisica Nucleare (Italy),
the Research Council of Norway, the
Ministry of Science and Technology of the Russian Federation, and the
Particle Physics and Astronomy Research Council (United Kingdom). 
Individuals have received support from the Swiss 
National Science Foundation, the A. P. Sloan Foundation, 
the Research Corporation,
and the Alexander von Humboldt Foundation.

\end{document}